\pdfoutput=1
\documentclass[sigplan,nonacm,10pt]{acmart}

\usepackage{xspace}
\usepackage{amsthm}
\usepackage{algorithm}
\usepackage{caption}
\usepackage{subcaption}
\usepackage[noend]{algpseudocode}
\usepackage{pgf}
\usepackage{bm}
\usepackage{bbm}
\usepackage{tikz}
\usepackage{comment}
\usepackage{float}

\usepackage{multirow}
\usepackage{enumitem}
\usepackage{multicol}
\usepackage{dsfont}
\usepackage[htt]{hyphenat}

\usepackage{titlesec}

\setlist{nosep}
\setlist{noitemsep}

\algrenewcommand\algorithmicrequire{\textbf{Arguments:}}
\algrenewcommand\algorithmicensure{\textbf{Returns:}}

\newcommand*\circled[1]{\tikz[baseline=(char.base)]{\node[shape=circle,fill,inner sep=0.1pt,minimum size=.35cm] (char) {\small{\textcolor{white}{#1}}};}}

\author{Jiacheng Yang}
\affiliation{%
  \institution{University of Toronto \&\\Vector Institute}
  \country{}
}
\author{Christina Giannoula}
\affiliation{%
  \institution{University of Toronto}
  \country{}
}
\author{Jun Wu}
\affiliation{%
  \institution{Amazon}
  \country{}
}
\author{Mostafa Elhoushi}
\affiliation{%
  \institution{Meta}
  \country{}
}
\author{James Gleeson}
\affiliation{%
  \institution{Samsung AI Centre Toronto}
  \country{}
}
\author{Gennady Pekhimenko}
\affiliation{%
  \institution{CentML \& University of Toronto \&\\Vector Institute}
  \country{}
}

\newcommand{\DefineEval}[2]{%
  \expandafter\newcommand\csname eval-#1\endcsname{#2}%
}
\newcommand{\Eval}[1]{\csname eval-#1\endcsname}

\DefineEval{MinuetE2EVSMinkowskiEngineA100Avg}{$1.63\times$\xspace}
\DefineEval{MinuetE2EVSMinkowskiEngineA100Min}{$1.36\times$\xspace}
\DefineEval{MinuetE2EVSMinkowskiEngineA100Max}{$1.87\times$\xspace}
\DefineEval{MinuetE2EVSTorchSparseA100Avg}{$1.83\times$\xspace}
\DefineEval{MinuetE2EVSTorchSparseA100Min}{$1.58\times$\xspace}
\DefineEval{MinuetE2EVSTorchSparseA100Max}{$2.08\times$\xspace}
\DefineEval{MinuetE2EVSA100}{$1.72\times$\xspace}
\DefineEval{MinuetE2EVSMinkowskiEngineRTX2070SuperAvg}{$1.80\times$\xspace}
\DefineEval{MinuetE2EVSMinkowskiEngineRTX2070SuperMin}{$1.43\times$\xspace}
\DefineEval{MinuetE2EVSMinkowskiEngineRTX2070SuperMax}{$2.19\times$\xspace}
\DefineEval{MinuetE2EVSTorchSparseRTX2070SuperAvg}{$1.68\times$\xspace}
\DefineEval{MinuetE2EVSTorchSparseRTX2070SuperMin}{$1.33\times$\xspace}
\DefineEval{MinuetE2EVSTorchSparseRTX2070SuperMax}{$2.10\times$\xspace}
\DefineEval{MinuetE2EVSRTX2070Super}{$1.74\times$\xspace}
\DefineEval{MinuetE2EVSMinkowskiEngineRTX2080TiAvg}{$1.78\times$\xspace}
\DefineEval{MinuetE2EVSMinkowskiEngineRTX2080TiMin}{$1.46\times$\xspace}
\DefineEval{MinuetE2EVSMinkowskiEngineRTX2080TiMax}{$2.12\times$\xspace}
\DefineEval{MinuetE2EVSTorchSparseRTX2080TiAvg}{$1.66\times$\xspace}
\DefineEval{MinuetE2EVSTorchSparseRTX2080TiMin}{$1.36\times$\xspace}
\DefineEval{MinuetE2EVSTorchSparseRTX2080TiMax}{$2.05\times$\xspace}
\DefineEval{MinuetE2EVSRTX2080Ti}{$1.72\times$\xspace}
\DefineEval{MinuetE2EVSMinkowskiEngineRTX3090Avg}{$1.76\times$\xspace}
\DefineEval{MinuetE2EVSMinkowskiEngineRTX3090Min}{$1.39\times$\xspace}
\DefineEval{MinuetE2EVSMinkowskiEngineRTX3090Max}{$2.18\times$\xspace}
\DefineEval{MinuetE2EVSTorchSparseRTX3090Avg}{$1.82\times$\xspace}
\DefineEval{MinuetE2EVSTorchSparseRTX3090Min}{$1.50\times$\xspace}
\DefineEval{MinuetE2EVSTorchSparseRTX3090Max}{$2.22\times$\xspace}
\DefineEval{MinuetE2EVSRTX3090}{$1.79\times$\xspace}
\DefineEval{MinuetE2EVSMinkowskiEngineAvg}{$1.74\times$\xspace}
\DefineEval{MinuetE2EVSMinkowskiEngineMin}{$1.36\times$\xspace}
\DefineEval{MinuetE2EVSMinkowskiEngineMax}{$2.19\times$\xspace}
\DefineEval{MinuetE2EVSTorchSparseAvg}{$1.74\times$\xspace}
\DefineEval{MinuetE2EVSTorchSparseMin}{$1.33\times$\xspace}
\DefineEval{MinuetE2EVSTorchSparseMax}{$2.22\times$\xspace}
\DefineEval{MinuetE2EAvg}{$1.74\times$\xspace}
\DefineEval{MinuetE2EMin}{$1.33\times$\xspace}
\DefineEval{MinuetE2EMax}{$2.22\times$\xspace}

\DefineEval{MinuetLayerwiseMinkowskiEngineRTX3090Avg}{$1.64\times$\xspace}
\DefineEval{MinuetLayerwiseMinkowskiEngineRTX3090Min}{$1.44\times$\xspace}
\DefineEval{MinuetLayerwiseMinkowskiEngineRTX3090Max}{$2.16\times$\xspace}
\DefineEval{MinuetLayerwiseTorchSparseRTX3090Avg}{$1.67\times$\xspace}
\DefineEval{MinuetLayerwiseTorchSparseRTX3090Min}{$1.28\times$\xspace}
\DefineEval{MinuetLayerwiseTorchSparseRTX3090Max}{$2.10\times$\xspace}
\DefineEval{MinuetLayerwiseRTX3090Avg}{$1.66\times$\xspace}
\DefineEval{MinuetLayerwiseRTX3090Min}{$1.28\times$\xspace}
\DefineEval{MinuetLayerwiseRTX3090Max}{$2.16\times$\xspace}

\DefineEval{MinuetMapVSMinkowskiEngineRTX2070Avg}{$14.8\times$\xspace}
\DefineEval{MinuetMapVSMinkowskiEngineRTX2070Min}{$11.3\times$\xspace}
\DefineEval{MinuetMapVSMinkowskiEngineRTX2070Max}{$23.8\times$\xspace}
\DefineEval{MinuetMapVSTorchSparseRTX2070Avg}{$18.2\times$\xspace}
\DefineEval{MinuetMapVSTorchSparseRTX2070Min}{$7.16\times$\xspace}
\DefineEval{MinuetMapVSTorchSparseRTX2070Max}{$28.3\times$\xspace}
\DefineEval{MinuetMapVSMinkowskiEngineRTX2080TIAvg}{$15.1\times$\xspace}
\DefineEval{MinuetMapVSMinkowskiEngineRTX2080TIMin}{$10.6\times$\xspace}
\DefineEval{MinuetMapVSMinkowskiEngineRTX2080TIMax}{$25.8\times$\xspace}
\DefineEval{MinuetMapVSTorchSparseRTX2080TIAvg}{$17.1\times$\xspace}
\DefineEval{MinuetMapVSTorchSparseRTX2080TIMin}{$4.72\times$\xspace}
\DefineEval{MinuetMapVSTorchSparseRTX2080TIMax}{$30.3\times$\xspace}
\DefineEval{MinuetMapVSMinkowskiEngineRTX3090Avg}{$19.2\times$\xspace}
\DefineEval{MinuetMapVSMinkowskiEngineRTX3090Min}{$13.9\times$\xspace}
\DefineEval{MinuetMapVSMinkowskiEngineRTX3090Max}{$26.8\times$\xspace}
\DefineEval{MinuetMapVSTorchSparseRTX3090Avg}{$13\times$\xspace}
\DefineEval{MinuetMapVSTorchSparseRTX3090Min}{$2.72\times$\xspace}
\DefineEval{MinuetMapVSTorchSparseRTX3090Max}{$24.6\times$\xspace}
\DefineEval{MinuetMapRTX2070Avg}{$16.4\times$\xspace}
\DefineEval{MinuetMapRTX2070Min}{$7.16\times$\xspace}
\DefineEval{MinuetMapRTX2070Max}{$28.3\times$\xspace}
\DefineEval{MinuetMapRTX2080TIAvg}{$16\times$\xspace}
\DefineEval{MinuetMapRTX2080TIMin}{$4.72\times$\xspace}
\DefineEval{MinuetMapRTX2080TIMax}{$30.3\times$\xspace}
\DefineEval{MinuetMapRTX3090Avg}{$15.8\times$\xspace}
\DefineEval{MinuetMapRTX3090Min}{$2.72\times$\xspace}
\DefineEval{MinuetMapRTX3090Max}{$26.8\times$\xspace}

\DefineEval{MinuetGMulSVSMinkowskiEngineRTX3090Avg}{$1.40\times$\xspace}
\DefineEval{MinuetGMulSVSMinkowskiEngineRTX3090Min}{$0.83\times$\xspace}
\DefineEval{MinuetGMulSVSMinkowskiEngineRTX3090Max}{$2.38\times$\xspace}
\DefineEval{MinuetGMulSVSTorchSparseRTX3090Avg}{$1.37\times$\xspace}
\DefineEval{MinuetGMulSVSTorchSparseRTX3090Min}{$1.22\times$\xspace}
\DefineEval{MinuetGMulSVSTorchSparseRTX3090Max}{$1.59\times$\xspace}
\DefineEval{MinuetGMulSRTX3090Avg}{$1.39\times$\xspace}
\DefineEval{MinuetGMulSRTX3090Min}{$0.83\times$\xspace}
\DefineEval{MinuetGMulSRTX3090Max}{$2.38\times$\xspace}

\DefineEval{TorchSparseGEMMPadding}{$11\%$}
\DefineEval{TorchSparseGEMMCount}{$11.1$}
\DefineEval{MinuetGEMMPadding}{$8.2\%$}
\DefineEval{MinuetGEMMCount}{$7.76$}

\DefineEval{MinuetSparsityMax}{$1.90\times$\xspace}
\DefineEval{MinuetSparsityAvg}{$1.68\times$\xspace}

\newcommand{\MinuetMainGPU}{RTX 3090\xspace}
\newcommand{\MinuetMainGPUCode}{rtx3090}
\DefineEval{MinuetMapAvg}{\Eval{MinuetMapRTX3090Avg}}
\DefineEval{MinuetMapMin}{\Eval{MinuetMapRTX3090Min}}
\DefineEval{MinuetMapMax}{\Eval{MinuetMapRTX3090Max}}

\DefineEval{MinuetLayerwiseAvg}{\Eval{MinuetLayerwiseRTX3090Avg}}
\DefineEval{MinuetLayerwiseMin}{\Eval{MinuetLayerwiseRTX3090Min}}
\DefineEval{MinuetLayerwiseMax}{\Eval{MinuetLayerwiseRTX3090Max}}

\DefineEval{MinuetLayerwiseVSMinkowskiEngineAvg}{\Eval{MinuetLayerwiseMinkowskiEngineRTX3090Avg}}
\DefineEval{MinuetLayerwiseVSMinkowskiEngineMin}{\Eval{MinuetLayerwiseMinkowskiEngineRTX3090Min}}
\DefineEval{MinuetLayerwiseVSMinkowskiEngineMax}{\Eval{MinuetLayerwiseMinkowskiEngineRTX3090Max}}

\DefineEval{MinuetLayerwiseVSTorchSparseAvg}{\Eval{MinuetLayerwiseTorchSparseRTX3090Avg}}
\DefineEval{MinuetLayerwiseVSTorchSparseMin}{\Eval{MinuetLayerwiseTorchSparseRTX3090Min}}
\DefineEval{MinuetLayerwiseVSTorchSparseMax}{\Eval{MinuetLayerwiseTorchSparseRTX3090Max}}

\DefineEval{MinuetMapVSMinkowskiEngineAvg}{\Eval{MinuetMapVSMinkowskiEngineRTX3090Avg}}
\DefineEval{MinuetMapVSMinkowskiEngineMin}{\Eval{MinuetMapVSMinkowskiEngineRTX3090Min}}
\DefineEval{MinuetMapVSMinkowskiEngineMax}{\Eval{MinuetMapVSMinkowskiEngineRTX3090Max}}

\DefineEval{MinuetMapVSTorchSparseAvg}{\Eval{MinuetMapVSTorchSparseRTX3090Avg}}
\DefineEval{MinuetMapVSTorchSparseMin}{\Eval{MinuetMapVSTorchSparseRTX3090Min}}
\DefineEval{MinuetMapVSTorchSparseMax}{\Eval{MinuetMapVSTorchSparseRTX3090Max}}

\DefineEval{MinuetGMulSVSMinkowskiEngineAvg}{\Eval{MinuetGMulSVSMinkowskiEngineRTX3090Avg}}
\DefineEval{MinuetGMulSVSMinkowskiEngineMin}{\Eval{MinuetGMulSVSMinkowskiEngineRTX3090Min}}
\DefineEval{MinuetGMulSVSMinkowskiEngineMax}{\Eval{MinuetGMulSVSMinkowskiEngineRTX3090Max}}

\DefineEval{MinuetGMulSVSTorchSparseAvg}{\Eval{MinuetGMulSVSTorchSparseRTX3090Avg}}
\DefineEval{MinuetGMulSVSTorchSparseMin}{\Eval{MinuetGMulSVSTorchSparseRTX3090Min}}
\DefineEval{MinuetGMulSVSTorchSparseMax}{\Eval{MinuetGMulSVSTorchSparseRTX3090Max}}

\DefineEval{MinuetGMulSAvg}{\Eval{MinuetGMulSRTX3090Avg}}
\DefineEval{MinuetGMulSMin}{\Eval{MinuetGMulSRTX3090Min}}
\DefineEval{MinuetGMulSMax}{\Eval{MinuetGMulSRTX3090Max}}

\newcommand{\system}{Minuet\xspace}
\newcommand{\systemRepo}{\color{blue}\url{https://github.com/UofT-EcoSystem/Minuet}}
\newcommand{\SpConv}{\emph{SC}\xspace}
\newcommand{\DNN}{\emph{DNN}\xspace}
\newcommand{\Map}{\emph{Map}\xspace}
\newcommand{\GMulS}{\emph{GMaS}\xspace}

\newcommand{\VRAR}{\emph{VR/AR}\xspace}
\newcommand{\LiDAR}{\emph{LiDAR}\xspace}

\title{\system: Accelerating 3D Sparse Convolutions on GPUs}

\begin{document}

\titlespacing*{\section}{0pt}{4pt}{2pt}
\titlespacing*{\subsection}{0pt}{2pt}{1pt}
\titlespacing*{\subsubsection}{0pt}{2pt}{2pt}

\begin{abstract}

Sparse Convolution (\SpConv) is widely used for processing 3D point clouds that are inherently sparse.
Different from dense convolution, \SpConv preserves the sparsity of the input point cloud by only allowing outputs to specific locations.
To efficiently compute \SpConv, prior \SpConv engines first use hash tables to build a kernel map that stores the necessary General Matrix Multiplication (GEMM) operations to be executed (\Map step), and then use a Gather-GEMM-Scatter process to execute these GEMM operations (\GMulS step). 
In this work, we analyze the shortcomings of prior state-of-the-art \SpConv engines, and propose \system, a novel memory-efficient \SpConv engine tailored for modern GPUs. 
\system proposes to (i) replace the hash tables used in the \Map step with a novel segmented sorting double-traversed binary search algorithm that highly utilizes the on-chip memory hierarchy of GPUs, (ii) use a lightweight scheme to autotune the tile size in the Gather and Scatter operations of the \GMulS step, such that to adapt the execution to the particular characteristics of each \SpConv layer, dataset, and GPU architecture, and (iii) employ a padding-efficient GEMM grouping approach that reduces both memory padding and kernel launching overheads. 
Our evaluations show that \system significantly outperforms prior \SpConv engines by on average \Eval{MinuetE2EAvg} (up to \Eval{MinuetE2EMax}) for end-to-end point cloud network executions. 
Our novel segmented sorting double-traversed binary search algorithm achieves superior speedups by \Eval{MinuetMapAvg} on average (up to \Eval{MinuetMapMax}) over prior \SpConv engines in the \Map step. The source code of \system is publicly available at \systemRepo.

\end{abstract}
\maketitle
\vspace{-8pt}
\section{Introduction}

Thanks to recent advances in 3D sensors, such as light detection and ranging (\LiDAR) sensors, 3D point clouds become increasingly accessible and widely used in many important applications, including virtual and augmented reality (\VRAR)~\cite{8814115}, photography~\cite{rs5126382}, drones~\cite{9213894}, robotics~\cite{KIM201838}, and autonomous vehicles~\cite{7989591,DBLP:journals/corr/abs-1904-01416}.
Similarly to popular deep neural networks (\DNN{s}), point cloud networks provide high efficiency and accuracy on a variety of vision tasks, such as 3D object detection~\cite{DBLP:journals/corr/abs-2110-02531,DBLP:journals/corr/abs-2012-11409,DBLP:journals/corr/abs-2109-02497} and segmentation~\cite{DBLP:journals/corr/abs-2007-16100,DBLP:journals/corr/abs-2003-06537,DBLP:journals/corr/abs-2110-02210}. 

\begin{figure}[t]
    \centering
    \includegraphics[width=0.99\linewidth]{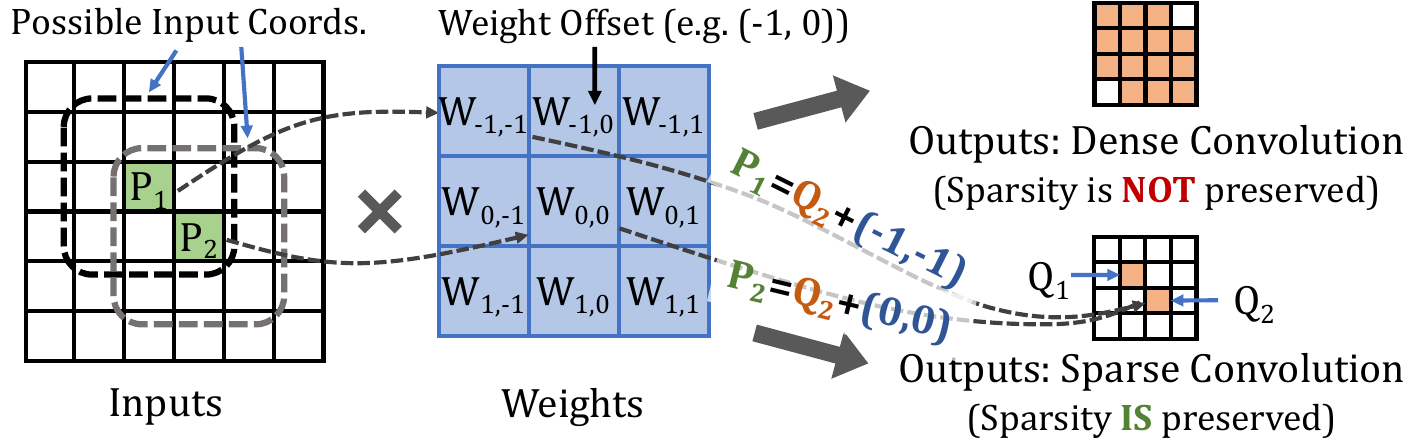}
    \caption{Dense convolution versus sparse convolution.}
    \label{fig:dense-and-sparse-conv}
\end{figure}

Different from 2D dense images, 3D point clouds describe 3D objects that are extremely sparse to their bounding space (usually less than 0.01\% within the bounding 3D volume~\cite{DBLP:journals/corr/abs-2110-07600}).
Therefore, researchers propose dedicated DNN-based algorithms~\cite{DBLP:journals/corr/abs-1904-08755,DBLP:journals/corr/abs-1907-03739,DBLP:journals/corr/QiSMG16,DBLP:journals/corr/QiYSG17,Chen_2023_CVPR,7353481,Liu_2023_CVPR,DBLP:journals/corr/abs-1712-05245} to efficiently process 3D point clouds by taking into consideration the sparse execution pattern.
Among these algorithms, Sparse Convolution (\SpConv) networks~\cite{3DSemanticSegmentationWithSubmanifoldSparseConvNet,DBLP:journals/corr/abs-1904-08755} achieve high accuracy, dominating performance, and wide applicability. 
As shown in Figure~\ref{fig:dense-and-sparse-conv}, unlike dense convolution where the sparsity is quickly diluted, \SpConv only allows the set of output points to specific locations that preserve the sparsity pattern exhibited in the input point cloud. 
Thus, to reduce the number of computations, for each output point, \SpConv needs to find the locations of the corresponding input point and the weight, which results in \emph{implicit} General Matrix Multiplications (GEMMs)~\cite{MLSYS2022_6512bd43,DBLP:journals/corr/abs-2110-07600}, i.e., the exact input feature and weight for each GEMM are implied by the sparsity pattern of the input point cloud.

To efficiently execute implicit GEMMs, prior works break the \SpConv execution into two steps: (1) the mapping step (\textbf{\Map}); and (2) the Gather-GEMM-Scatter step (\textbf{\GMulS}). 
In the \Map step, \SpConv builds a \emph{kernel map} that stores the necessary GEMM operations needed to be performed, i.e., the indices of the weights and the input/output feature vectors.
In the \GMulS step, \SpConv executes each necessary GEMM operation in the kernel map to transform the input feature vectors into the output feature vectors.
To build the kernel map in the \Map step, prior \SpConv engines \cite{MLSYS2022_6512bd43,DBLP:journals/corr/abs-2110-07600,spconv2022} create \emph{queries} with all possible input coordinates by enumerating the additions of each output coordinate with each weight offset.
Then, they check the existence of non-zero input data points by executing the queries to a \emph{hash table}, that stores the coordinates of the actual non-zero input data points. 
In the \GMulS step, prior \SpConv engines \cite{MLSYS2022_6512bd43} use an input buffer array and an output buffer array to \emph{continuously} store the operands of the GEMM operations, i.e., the values of the input and the output feature vectors. This approach enables prior \SpConv engines to leverage highly-performant GPU GEMM libraries (e.g., cuBLAS \cite{cublas}) that require GEMM operands to be continuous in memory.
To do so, they first broadcast the input feature vectors to the input buffer array with a \emph{Gather} operation, then execute GEMM operations to create partial results for output feature vectors which are stored in the output buffer array, and finally merge (sum-reduce) the partial results to assemble the final output feature vectors with a \emph{Scatter} operation. 

In this work, we characterize existing \SpConv engines~\cite{DBLP:journals/corr/abs-1904-08755,MLSYS2022_6512bd43,spconv2022} using various point cloud networks, real datasets, and GPU architectures, and find that they suffer from three key shortcomings. 
First, they use a hash table (e.g., cuckoo hash table~\cite{10.1145/1618452.1618500}) to build the kernel map, which stores the necessary GEMM operations. 
However, executing a large number of \emph{queries} in the hash table incurs irregular data accesses, most of which are served by the expensive GPU global memory, thus causing high data access costs.
Second, state-of-the-art \SpConv engines process multiple input/output feature channels of \SpConv in \emph{tiles}, as a chunk of consecutive feature channels, in Gather and Scatter operations to improve GPU memory throughput. 
However, we observe that they always employ a single \emph{fixed} tile size, which suffers from sub-optimal performance. 
In Figure~\ref{fig:gs_tile_size}, we demonstrate that the best-performing tile size depends on the characteristics of each particular \SpConv layer of the point cloud network, the real dataset, and the GPU architecture. 
Third, in the \GMulS step, prior \SpConv engines execute GEMM operations corresponding to multiple weight offsets in a \emph{batched} scheme: they group multiple GEMM operations together by padding with zero values the GEMM operands, i.e., they provide the same sizes among all GEMM operands and execute them as a single batched GEMM kernel. 
This way they minimize GPU kernel launch overheads and improve GPU hardware utilization \cite{MLSYS2022_6512bd43}.
However, we find that prior \SpConv engines group GEMM operations in the \GMulS step following the order induced by the \Map step, i.e., the order of the weight offsets. This approach causes high padding overheads (Section~\ref{sec:motivation}), i.e., a large number of zero values are added, thus incurring many redundant data accesses and computations.

To tackle the aforementioned issues, we propose \system, a novel memory-efficient \SpConv engine tailored for modern GPUs. 
\system highly utilizes the on-chip memory hierarchy of GPUs, adapts \SpConv execution to the characteristics of the input dataset and GPU architecture, and reduces unnecessary data accesses and computations. 
In the \Map step, we challenge the prevailing notion that the hash table-based search performs superiorly compared to binary search on GPUs~\cite{ALCANTARA201239,10.1145/1618452.1618500}, and propose an innovative binary search-based algorithm tailored for building the kernel maps in the \Map step of \SpConv on modern GPU architectures. We leverage the key observation that when executing \emph{sorted} queries, binary search achieves high system efficiency, leveraging data locality across consecutive \emph{sorted} queries, and propose the \emph{segmented sorting double-traversed} binary search algorithm. 
Our proposed algorithm for \SpConv can achieve a similar theoretical computational complexity with the hash table-based search (Section \ref{sec:bs-complexity}) and provides significantly higher memory efficiency, improving the hit ratio in the on-chip caches of GPUs (Figure \ref{fig:eval-map-query-time}b). 
In the \GMulS execution step, \system provides two optimizations. 
First, we on-the-fly tune the tile size used to process multiple input/output feature channels at \emph{each} Gather and Scatter operations. This key technique enables \system to adapt the \SpConv execution to the particular characteristics of each \SpConv layer in point cloud networks, real dataset, and GPU architecture, thus providing high system performance in Gather and Scatter operations. 
Second, \system integrates a padding-efficient GEMM grouping strategy, which first reorders GEMM operations based on the sizes of input/output feature vectors, and then groups GEMM operations into batched GEMM kernel launches. 
This way \system optimizes both (i) the amount of padding with zero values, thus minimizing unnecessary data accesses and computations to useless data in GEMM kernels, and (ii) the GEMM kernel launch overheads.

We extensively evaluate \system using a wide variety of 3D point cloud networks, real datasets, and GPU architectures, and demonstrate that \system significantly outperforms prior works. 
Compared to state-of-the-art \SpConv engines, \system improves the end-to-end performance by \Eval{MinuetE2EAvg} on average (up to \Eval{MinuetE2EMax}), and  achieves superior speedups over prior \SpConv engines in the \Map step by on average \Eval{MinuetMapAvg} (up to \Eval{MinuetMapMax}), and by on average  \Eval{MinuetGMulSAvg} (up to \Eval{MinuetGMulSMax}) in the \GMulS step.

Overall, this paper makes the following contributions:
\begin{itemize}[noitemsep,topsep=0pt,leftmargin=8pt]
   \item We investigate the shortcomings of existing \SpConv engines, and propose \system, a memory-efficient engine to accelerate \SpConv executions on modern GPU architectures.
   \item We propose a novel segmented sorting double-traversed binary search algorithm to build kernel maps in \SpConv. Our proposed algorithm highly utilizes the on-chip memory hierarchy of GPUs. 
   We also dynamically select the best-performing tile size in Gather and Scatter operations, and reorder GEMM operations before grouping them to minimize unnecessary data accesses and computations.
   \item We evaluate \system using a wide range of real datasets, sparse 3D networks, and GPU architectures, and show that it significantly outperforms prior works both in layerwise and end-to-end execution. \system also provides superior speedups in the \Map step of \SpConv. 
   We open-source \system at \systemRepo.
\end{itemize}
\begin{figure*}
    \centering
    \includegraphics[width=\linewidth]{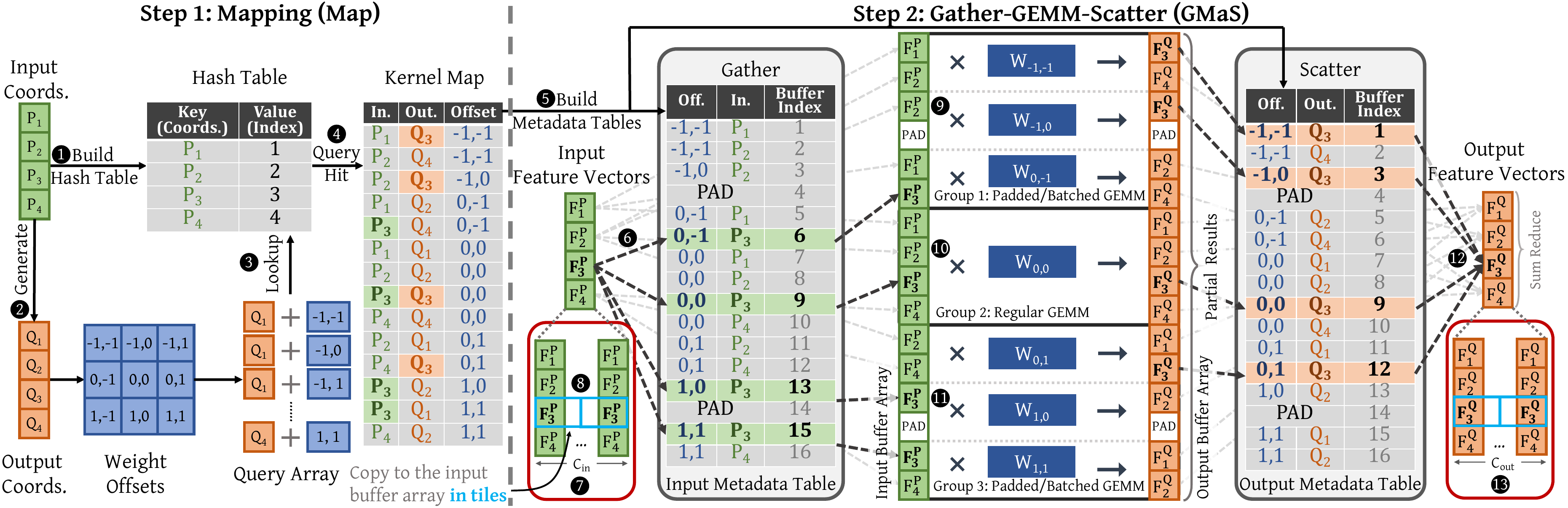}
    \caption{The \SpConv execution can be broken down into two steps. For simplicity, we use 2D coordinates for illustration.} 
    \label{fig:spconv}
\end{figure*}

\section{Sparse Convolution (\SpConv)}

\subsection{\SpConv Definition}\label{sec:background-definition}

A 3D object is inherently sparse in nature, i.e., it does not completely fill the 3D space it occupies, thus resulting in a spatially sparse structure.
Point cloud is a widely applied sparse format that is used to effectively represent a 3D object, thanks to its simplicity and accuracy \cite{DBLP:journals/corr/abs-1912-12033,DBLP:journals/corr/QiSMG16,8814115,rs5126382,9213894,KIM201838,7989591}.
Specifically, a point cloud only stores the non-zero points of a 3D object as an unordered set of points $\mathcal{P} = \{\mathbf{p_i}\}$ and its corresponding set of feature vectors $\{\mathbf{F^\mathcal{P}_i}\}$.
Each point $\mathbf{p_i}$ is a 3D coordinate that represents one non-zero point of the 3D object, and each feature vector $\mathbf{F^\mathcal{P}_i}$ of size $C$ stores the corresponding $C$ feature channels (e.g., $C = 3$ for RGB colors) of the $i$-th point $\mathbf{p_i}$. 
Thus, the feature vectors of a point cloud with $N$ points can be stored in an $N \times C$ feature matrix $\mathbf{F}^\mathcal{P}$.

Sparse Convolution (\SpConv) takes as input a 3D object represented as a point cloud, and its output is also a point cloud that preserves the sparsity pattern of the input point cloud.
This is achieved by only allowing the computations of output feature vectors on specific output coordinates that are generated based on the input coordinates. 
More formally, \SpConv uses the following formula to generate the output coordinates $\mathcal{Q}$ based on the input coordinates $\mathcal{P}$ with a stride parameter $s$:
\begin{equation}\label{eq:output-generate}\textstyle
    \mathcal{Q} = \left\{\left(\left\lfloor\frac{x}{s}\right\rfloor \times s, \left\lfloor\frac{y}{s}\right\rfloor \times s, \left\lfloor\frac{z}{s}\right\rfloor \times s\right) | (x, y, z) \in \mathcal{P}\right\}
\end{equation}
To keep output coordinates unique, duplicates among the output coordinates are eliminated.
Intuitively, the output coordinates $\mathcal{Q}$ are downsampled from the input coordinates $\mathcal{P}$ and the stride parameter $s$ specifies the granularity of the downsampling. 
Note that if the stride $s$ is equal to $1$, the output coordinates will be the same as the input coordinates, i.e., $\mathcal{Q} = \mathcal{P}$. 
Then, the output feature vector $\mathbf{F^\mathcal{Q}_i}$ of the $i$-th output coordinate $\mathbf{q_i}$ is computed on every weight offset $\boldsymbol{\delta_k}$ and every input coordinate $\mathbf{p_j}$, when the condition $\mathbf{p_j} = \mathbf{q_i} + \boldsymbol{\delta_k}$ holds, which is formalized as follows:
\begin{equation}\label{eq:sparse-conv-definition}\textstyle
    \mathbf{F^\mathcal{Q}_i} = \sum_{\boldsymbol{\delta}_\mathbf{k} \in \boldsymbol{\Delta}(K, \mathbf{s})} \sum_{p_j \in \mathcal{P}} \mathds{1}_{\mathbf{p_j} = \mathbf{q_i} + \boldsymbol{\delta}_\mathbf{k}} \mathbf{F^\mathcal{P}_j} \mathbf{W}_{\boldsymbol{\delta}_\mathbf{k}}\quad(\mathbf{q_i} \in \mathcal{Q})
\end{equation}
where $\boldsymbol{\Delta}(K, s)$ stands for the set of \emph{weight offsets} with a kernel size $K$ and a stride $s$ (e.g., $\boldsymbol{\Delta}(5, 2) = \{-4, -2, 0, 2, 4\}^3$), $\boldsymbol{\delta}_\mathbf{k}$ for the $k$-th weight offset (e.g. $\boldsymbol{\delta}_\mathbf{1} = (-4, -4, -4) \in \boldsymbol{\Delta}(5, 2)$), $\mathbf{F^\mathcal{P}_j}$ for the feature vector of the input coordinate $\mathbf{p_j}$, $\mathbf{W}_{\boldsymbol{\delta}_\mathbf{k}}$ for the weight corresponding to the weight offset $\boldsymbol{\delta}_\mathbf{k}$, and $\mathds{1}_{\mathbf{p_j} = \mathbf{q_i} + \boldsymbol{\delta}_\mathbf{k}}$ for the indicator function on the condition $\mathbf{p_j} = \mathbf{q_i} + \boldsymbol{\delta}_\mathbf{k}$. 

\subsection{Execution Steps of \SpConv}\label{sec:background-execsteps}

To effectively perform \SpConv execution, existing \SpConv frameworks construct an input-output map $\mathcal{M} = \{(\mathbf{p_j}, \mathbf{q_i}, \boldsymbol{\delta}_\mathbf{k})\}$, named as \textbf{kernel map}. Each kernel map entry represents a General Matrix Multiplication (GEMM) operation in Equation~\ref{eq:sparse-conv-definition}: 
\begin{equation}\label{eq:kernel-map-definition}\textstyle
    \forall (\mathbf{p_j}, \mathbf{q_i}, \boldsymbol{\delta}_\mathbf{k}) \in \mathcal{M}\quad\mathbf{F^\mathcal{Q}_i} \gets \mathbf{F^\mathcal{Q}_i} + \mathbf{F^\mathcal{P}_j} \mathbf{W}_{\boldsymbol{\delta}_\mathbf{k}}
\end{equation}
By traversing the kernel map, they perform only the necessary GEMM operations to compute the output feature vector for each output coordinates.

Figure~\ref{fig:spconv} describes the \SpConv execution, which can be broken down into two steps.

\noindent\textbf{Step 1: Mapping Step (\Map).}
To build the kernel map, existing implementations~\cite{DBLP:journals/corr/abs-1904-08755,spconv2022,MLSYS2022_6512bd43} first create a hash table \circled{1} where the input coordinates $\mathbf{p_j}$ are the keys and their indices $j$ are the values. 
Then, they  generate the output coordinates $\mathcal{Q}$ \circled{2} according to Equation \ref{eq:output-generate}, and create \emph{queries} $\mathbf{q_i} + \boldsymbol{\delta}_\mathbf{k}$ for each output coordinate $\mathbf{q_i}$ and each weight offset $\boldsymbol{\delta}_\mathbf{k}$ as the candidate input coordinates. 
Next, they perform lookup in the hash table  \circled{3} to check if each candidate input coordinate exists as an input coordinate $\mathbf{p_j}$.  
If such input coordinate $\mathbf{p_j}$ is found in the hash table, i.e., $\mathbf{p_j} = \mathbf{q_i} + \boldsymbol{\delta}_\mathbf{k}$, 
then a new entry $(\mathbf{p_j}, \mathbf{q_i}, \boldsymbol{\delta}_\mathbf{k})$ \circled{4}  is added to the kernel map, which corresponds to a necessary GEMM operation that needs to be performed to get the output feature vectors (Equation \ref{eq:kernel-map-definition}).

\noindent\textbf{Step 2: Gather-GEMM-Scatter Step (\GMulS).}
Executing the GEMM operation for \emph{each} entry in the kernel map results in a large number of small GEMM kernels, which incurs \emph{immense} kernel launch overheads in GPUs. 
Thus, existing frameworks~\cite{spconv2022,MLSYS2022_6512bd43} perform all GEMM operations associated with each weight with a \GMulS step (Figure~\ref{fig:spconv} right).

Specifically, for each weight, existing \SpConv engines \textbf{\emph{Gather}} the corresponding input feature vectors and store them consecutively to an input buffer array. 
To do so, they  build a metadata table \circled{5} which stores the positions that each input feature vector needs to be stored within the input buffer array.
For example, in Figure~\ref{fig:spconv}, 
the feature vector corresponding to the input coordinate $\mathbf{p_3}$ (i.e., $\mathbf{F}^\mathcal{P}_3$) is associated with $4$ entries in the kernel map, corresponding to the $4$ highlighted lines in the input metadata table. The input feature vector $\mathbf{F^\mathcal{P}_3}$ will be then copied to the corresponding positions in the input buffer array \circled{6} , i.e., the $6$-th, $9$-th, $13$-th, and $15$-th entries of the input buffer array. 
Note that each input feature vector has $C_\text{in}$ feature channels \circled{7}.
Thus, to maximize the GPU memory throughput,  state-of-the-art work~\cite{MLSYS2022_6512bd43} processes the feature channels of each input feature vector \emph{in tiles} \circled{8}, where each tile contains a \emph{fixed} size (typically $128$ bytes) of consecutive feature channels.
Specifically, each GPU thread loads one tile of the input feature vector to the on-chip register files, then accesses the input metadata table to find the corresponding buffer index of the input buffer array for \emph{each} tile, and finally copies the tile to the input buffer array. 

Then, \SpConv implementations perform one \textbf{GEMM} operation for each weight to create partial results for the output feature vectors. 
State-of-the-art \SpConv engines~\cite{MLSYS2022_6512bd43} group multiple GEMM operations corresponding to multiple weights, such that to be performed as a single batched GEMM kernel to minimize kernel launch overheads and leverage existing GPU GEMM libraries~\cite{cublas}. 
In the example of Figure~\ref{fig:spconv}, the 
GEMM operations are merged in $3$ GEMM groups \circled{9}, \circled{10}, and \circled{11}, and executed as $3$ GEMM kernel launches. 
To do so, padding with zero values is needed, and thus the buffer indices stored in the input metadata table of Gather operation are updated, accordingly. 
For instance, to perform the batched GEMM kernel launch of Group $1$ in Figure~\ref{fig:spconv} \circled{9}, \SpConv implementations pad with zero value feature channels in the buffer index $4$ of the input buffer array, such that the associated feature vectors of weight offset $(-1, 0)$ have the same sizes with that of the other two weight offsets of the same GEMM group, i.e., the weight offsets $(-1, -1)$ and $(0, -1)$.

Finally, the partial results produced by the batched GEMM operations in the output buffer array are  aggregated and reduced with a \textbf{Scatter} operation \circled{12} to obtain the output feature vectors. 
Similar to the Gather operation, in the Scatter operation, 
(i) an output metadata table is built to store the positions of the partial results in the output buffer array for each output feature vector, 
where the final values of the output feature vector is produced by merging these partial results.
(ii) padding is added in the indices stored in the output metadata table due to GEMM grouping on batched GEMM operations, 
and (iii) the feature channels of each output feature vector are processed \emph{in tiles} of the same sizes (typically $128$ bytes) \circled{13}: each GPU thread  loads one \emph{tile} of partial results in the output buffer array to the on-chip register files, then accesses the output metadata table to find the corresponding buffer index of the output buffer array for \emph{each} tile, and finally merges (sum-reduces) the tile to the output feature vectors. 

\section{Existing \SpConv Engines}\label{sec:motivation}

There are a few prior works~\cite{DBLP:journals/corr/abs-2110-00511,spconv2022,MLSYS2022_6512bd43} that optimize \SpConv on GPUs. 
MinkowskiEngine~\cite{DBLP:journals/corr/abs-1904-08755} is the first open-source library that efficiently implements \SpConv on GPUs, and is specifically optimized for \SpConv layers with small feature channel sizes. 
SpConv~\cite{spconv2022} improves the \SpConv execution by leveraging data locality in GEMM operations. TorchSparse~\cite{MLSYS2022_6512bd43} 
uses a single Gather and a single Scatter operation for \emph{all} weight offsets, and groups GEMM operations by performing zero padding in GEMM's operands, thus reducing kernel launch overheads and increasing GPU hardware utilization.

We comprehensively examine existing \SpConv engines~\cite{DBLP:journals/corr/abs-1904-08755,MLSYS2022_6512bd43,spconv2022} on a wide variety of real-world point cloud data, and find that prior approaches suffer from three shortcomings.

\noindent\textbf{Shortcoming \#1: Expensive Data Accesses in the \Map Step.}
To build the kernel map, prior works employ a hash table (Figure~\ref{fig:spconv} left) to store the coordinates of input point data. Then, they query the hash table for each output coordinate and weight offset to check the existence of the corresponding input coordinate. We observe that using a hash table to build the kernel map incurs a large number of irregular memory accesses that are served by GPU global memory,  thus resulting in low system performance.
We conclude that prior \SpConv frameworks do not effectively utilize the deep on-chip memory hierarchy of modern GPUs.

\begin{figure}[t]
    \centering
    \resizebox{\linewidth}{!}{\input{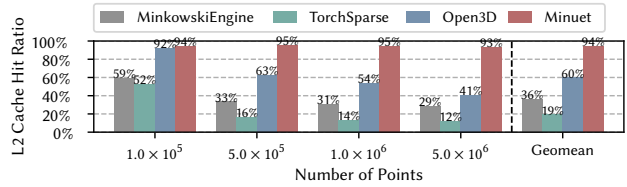}}
    \caption{L2 cache hit ratio in building kernel maps of the \Map step on \MinuetMainGPU for various kernel map building implementations.}
    \label{fig:l2_hit_rate_motivation}
\end{figure}

Figure~\ref{fig:l2_hit_rate_motivation} evaluates the hit ratio in the last level cache of GPUs, when building the kernel map in \SpConv execution using the hash table implementation of TorchSparse~\cite{MLSYS2022_6512bd43}, the hash table implementation of MinkowskiEngine~\cite{DBLP:journals/corr/abs-1904-08755}, the state-of-the-art 3D spatial hash table implementation on GPUs, i.e., Open3D~\cite{DBLP:journals/corr/abs-2110-00511}, and the implementation of our work \system. 
We find that prior state-of-the-art approaches achieve very low L2 cache hit ratio, on average $36\%$ and $19\%$ for the MinkowskiEngine~\cite{DBLP:journals/corr/abs-1904-08755} and TorchSparse~\cite{MLSYS2022_6512bd43}, respectively, due to random memory  access patterns incurred in their hash table-based implementations. As the number of input points increases, we also observe that hash table-based implementations have even lower cache hit ratios.
Even if the best-performing state-of-the-art hash table implementation, i.e., Open3D~\cite{DBLP:journals/corr/abs-2110-00511}, was used in \SpConv to build the kernel map, the L2 cache hit ratio would only be $41\%$ for a large number of data points, i.e., $5\times10^6$ points. 
Instead, we follow a fundamentally different approach in \system by employing a binary search-based algorithmic scheme to build the kernel map. We design \system to highly utilize the on-chip memory hierarchy of GPUs, thus providing high memory efficiency in the kernel map building of \SpConv.
Figure~\ref{fig:l2_hit_rate_motivation} demonstrates that \system achieves at least $93\%$ L2 cache hit ratio (even for a large number of data points), thus significantly improving the performance in the \Map step (See also Figure~\ref{fig:eval-map-query-time}).

\noindent\textbf{Shortcoming \#2: Sub-Optimal Performance in Gather and Scatter Operations.}
\begin{figure}[t]
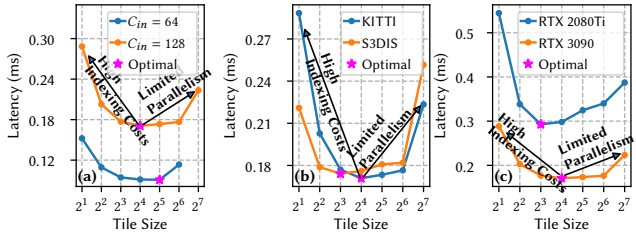

    \centering
    \begin{subfigure}[b]{0.33\linewidth}
        \centering
        \resizebox{\linewidth}{!}{\input{figures/figure10a-gather-scatter-diff-parameters-diff-layer.rtx3090.pgf}}
    \end{subfigure}
    \begin{subfigure}[b]{0.33\linewidth}
        \centering
        \resizebox{\linewidth}{!}{\input{figures/figure10b-gather-scatter-diff-parameters-diff-dataset.\MinuetMainGPUCode.pgf}}
    \end{subfigure}%
    \begin{subfigure}[b]{0.33\linewidth}
        \centering
        \resizebox{\linewidth}{!}{\input{figures/figure10c-gather-scatter-diff-parameters-diff-gpu.pgf}}
    \end{subfigure}%
    \caption{The performance of a Gather operation, when varying the (a) input channel size, (b) real dataset, and (c) GPU architecture.}
    \label{fig:gs_tile_size}
\end{figure}
In Gather and Scatter operations, prior works~\cite{MLSYS2022_6512bd43} use a fixed tile size to process the multiple input/output feature channels.
However, we observe that the best-performing tile size depends on the configuration of the \SpConv layer, the real dataset and the GPU architecture. Figure~\ref{fig:gs_tile_size} presents the latency in Gather operation for various tile sizes, when varying the (a) input channel size (layer configuration), (b) real dataset, and (c) GPU architecture.
On the one hand, using a small tile size to process input/output feature vectors results in many tiles corresponding to the \emph{same} buffer index of the input/output metadata tables. 
With this approach, \emph{each} entry in the metadata table is accessed multiple times,
thus resulting in \emph{high indexing costs} in metadata tables with significant performance overheads.
On the other hand, using a large tile size leads to fewer tiles to be parallelized, and thus results in \emph{limited execution parallelism}, causing hardware resource underutilization.
Moreover, the best-performing tile size depends on the channel size, input dataset, and GPU architecture.
Prior works overlook the aforementioned trade-off by using a single \emph{fixed} tile size in \SpConv execution,
and thus incur either high metadata indexing costs or low execution parallelism, which results in sub-optimal performance. 
Instead, \system provides a lightweight adaptive policy that dynamically autotunes the tile size based on the characteristics of each layer, real dataset, and GPU architecture, and thus achieves near optimal performance in Gather and Scatter operations.

\begin{figure}[t]
    \centering
    \includegraphics[width=\linewidth]{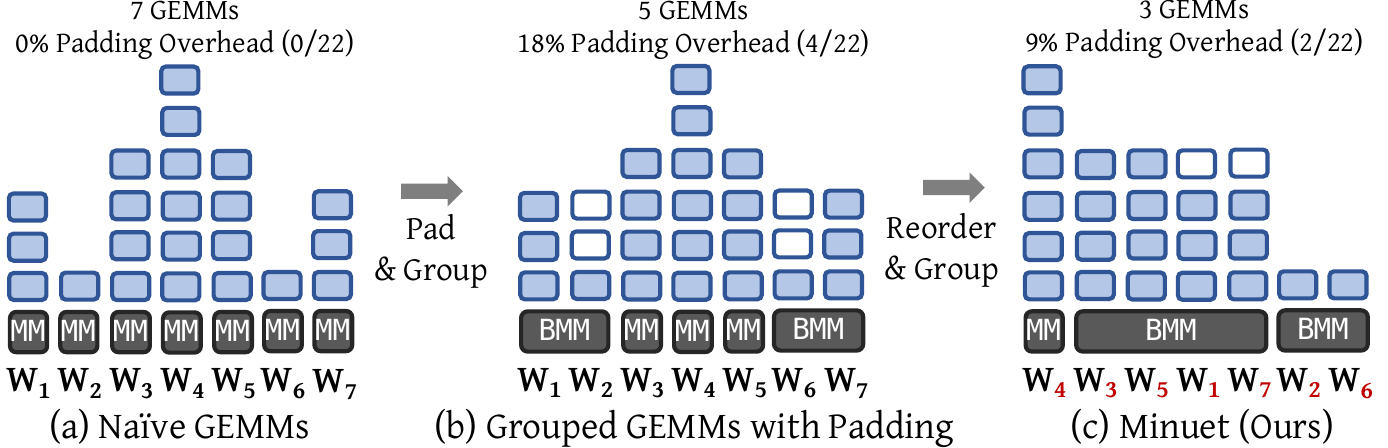}
    \caption{Various approaches to execute GEMM operations in \SpConv, where one blue and white squares denote one actual input feature vector and one zero-padded feature vector, respectively. Assuming $x$ and $y$ are the number of padded feature vectors and actual input feature vectors, respectively, the padding overhead is defined as $(x / y)$.}
    \label{fig:gemm-reorder}
\end{figure}

\noindent\textbf{Shortcoming \#3: High Padding Overhead in GEMM Operations.}
Using a na\"ive approach to execute each small GEMM kernel separately for each weight offset in the \GMulS step (Figure~\ref{fig:gemm-reorder}a) incurs excessive kernel launch overheads in GPUs, as explained in prior works~\cite{MLSYS2022_6512bd43,DBLP:journals/corr/abs-2007-01277,10.1109/GreenCom-CPSCom.2010.102}.
Thus, prior works~\cite{MLSYS2022_6512bd43} propose a batched GEMM approach, shown in Figure~\ref{fig:gemm-reorder}b: they group multiple GEMM operations together, padding with zero values the corresponding matrices (e.g., see the $2$-nd and $6$-th column in Figure~\ref{fig:gemm-reorder}b) of adjacent GEMM operations to have the same height, and launch one \emph{single} batched GEMM kernel for multiple weights. 
This grouping approach improves hardware utilization and kernel launch overheads in GEMM operations. However, we observe that this approach incurs high padding overhead, since prior works group GEMM operations in the \GMulS step following the order induced by the \Map step, i.e., the order in which weight offsets are processed in the \Map step. 
As a result, adjacent GEMM operations with that ordering might have a large difference in their sizes, causing a larger amount of padding with zero values, which in turn results to redundant data accesses and computations with zero (useless) values.

Instead, we argue that reordering the weights before grouping the GEMM operations can reduce the amount of padding with zero values, and provide a better GEMM grouping with lower padding overhead. For instance, Figure~\ref{fig:gemm-reorder}b shows that grouping GEMM operations in the order induced by the \Map step incurs $18\%$ padding overhead and launches $5$ GEMM kernels. However, if we first reorder weights carefully, and then group the GEMMs into batched GEMM kernel launches, we can provide only $9\%$ padding overhead and launch only $3$ GEMMs, as shown Figure \ref{fig:gemm-reorder}c.
To this end, we design \system to implement a lightweight GEMM reordering group policy that reduces the padding overhead and also provides a small number of GEMM kernel launches. 
Our evaluations show that prior state-of-the-art \SpConv work~\cite{MLSYS2022_6512bd43} incurs on average \Eval{TorchSparseGEMMPadding} padding overhead and executes on average \Eval{TorchSparseGEMMCount} GEMM kernels, while \system has \Eval{MinuetGEMMPadding} padding overhead and executes \Eval{MinuetGEMMCount} GEMM kernels.

\begin{figure*}
    \centering
    \includegraphics[width=\linewidth]{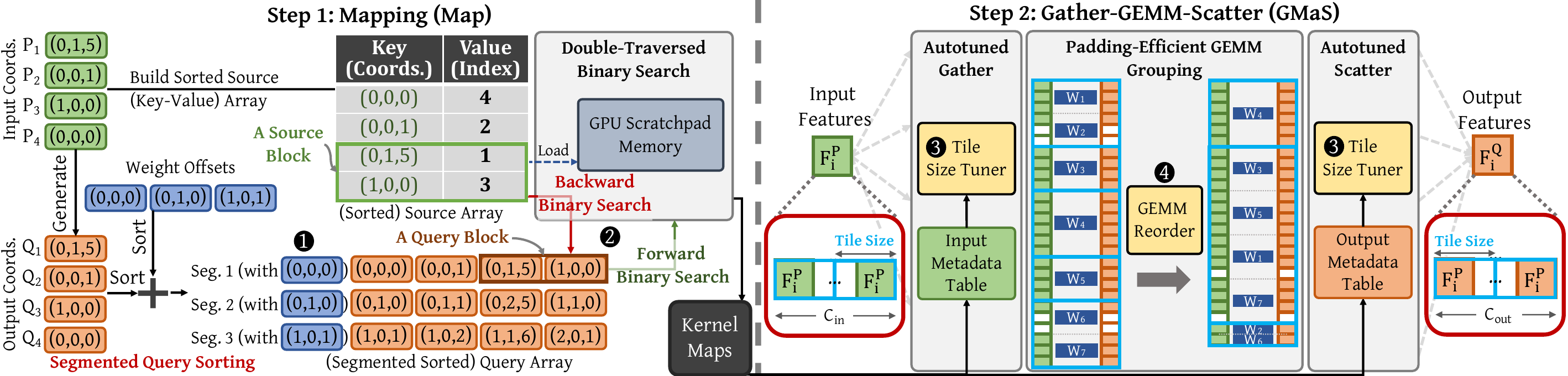}
    \caption{High-level overview of \system.}
    \label{fig:overview}
\end{figure*}

\section{\system: Overview}

\system is a novel high-performance \SpConv engine tailored for modern GPUs.
\system highly utilizes the on-chip memory hierarchy of GPUs, eliminates unnecessary data access and computations, and effectively adapts to both the data distribution of each particular input dataset and the characteristics of the GPU architecture used. 

Unlike prior \SpConv engines \cite{spconv2022,MLSYS2022_6512bd43,DBLP:journals/corr/abs-1904-08755} that use a hash table (e.g., cuckoo hash tables~\cite{10.1145/1618452.1618500}) for building kernel maps, in this work we argue that using a sorted key-value array and \emph{binary search} is a more efficient alternative solution than hash tables on GPUs.
Even though the na\"ive binary search in a sorted array has worse theoretical computational complexity than hash tables and does not effectively leverage the on-chip memory hierarchy of GPUs~\cite{ALCANTARA201239,10.1145/1618452.1618500}, we challenge these two understandings by proposing a novel binary search-based algorithm tailored for building kernel maps in \SpConv{s} on GPUs. 
Figure \ref{fig:overview} presents the high-level overview of \system.
Overall, we propose four key ideas that accelerate both the \Map and the \GMulS step in \SpConv execution.

\noindent\textbf{1. Segmented Query Sorting:} 
We sort the output coordinates and weight offsets separately, and create a query array in the \Map step, organizing the queries to \emph{sorted segments} \circled{1}: each \emph{query segment} is sorted and consists of the queries corresponding to all output coordinates associated with the \emph{same} weight offset. 
Then, for each query segment, we execute binary search queries in a sorted array that stores the input coordinates and their indices, named as \emph{source array}.
This way when we perform binary search lookups by iterating over consecutive sorted queries, we leverage temporal data locality: consecutive queries of the same sorted query segment have similar data access patterns in the source array, i.e., they access same elements in the source array with a high probability.
Segmented query sorting both minimizes the sorting overheads and improves data locality in on-chip caches of GPUs within the source array, thus accelerating performance to build kernel maps.

\noindent\textbf{2. Double-Traversed Binary Search:}
We split the source array into small disjoint \emph{source blocks}. For each source block, we perform a \emph{backward} binary search \circled{2} to each query segment to find out all possible queries corresponding to that source block, and these queries are organized as a \emph{query block}. 
Then, we load each \emph{source block} in the GPU scratchpad memory, and process all queries in the associated \emph{query block} by executing a \emph{forward} binary search within the \emph{source block}.
For each query, the proposed double-traversed binary search algorithm reduces the search range, since only a subset of the source array elements need to be compared, thus decreasing the number of computations (comparisons) performed, and provides low data access costs, by highly utilizing the on-chip memory hierarchy on GPUs.

\noindent\textbf{3. Autotuned Gather/Scatter:}
We design a tile size tuner that autotunes \circled{3} the tile size in Gather and Scatter operations for each \SpConv layer. First, we sample a few input point clouds from the dataset and create the corresponding input and output metadata table entries for these samples. Then, we evaluate the latency for all possible tile sizes and find the best-performing tile size for Gather and Scatter operation. Finally, we process all input point clouds from the dataset using the selected best-performing tile size. By autotuning the tile size at each \SpConv layer, \system achieves low metadata indexing costs and high execution parallelism in Gather and Scatter operations, and effectively adapts itself to the characteristics of each layer, dataset, and GPU architecture. 

\noindent\textbf{4. Padding-Efficient GEMM Grouping:} We re-order the GEMM operations \circled{4} based on the sorted sizes of their corresponding input and output feature vectors.
Then, we group adjacent GEMM operations (to be executed as a single GEMM kernel) associated with feature vectors of same or very similar sizes, which allows us to minimize the amount of zero paddings. With this key technique, we reduce redundant data accesses and computations with zero values, and minimize the number of GEMM kernels executed, thus enabling low kernel launch overheads on GPUs. 

\section{\system: Design Details}

\subsection{Optimizing the \Map Step}\label{sec:opt-map-step}

To build the kernel map, the coordinates of the non-zero data points in the input point cloud, i.e., the input coordinates $\{\mathbf{p_j}\}$, are stored to an array to be searched from (henceforth referred to as \emph{source array}), the size of which is denoted by $|\mathcal{P}|$. 
\SpConv creates an array of queries (henceforth referred to as \emph{query array}) that stores all possible input coordinates $\{\mathbf{q_i} + \boldsymbol{\delta}_\mathbf{k}\}$, where $\{\mathbf{q_i}\}$ and $\{\boldsymbol{\delta}_\mathbf{k}\}$ are the output coordinates and weight offsets, respectively.
Then, \SpConv executes each query to check whether the query exists as an input coordinate in the source array. 
Assuming an \SpConv layer with kernel size $K$ and $|\mathcal{Q}|$ output coordinates, the size of the query array is $K^3|\mathcal{Q}|$. 

\noindent\textbf{Key Observation.} 
\emph{Sorting the queries and executing them via binary search comprises many common elements in the search paths between adjacent queries.}

\begin{figure}[t]
    \centering
    \includegraphics[width=0.85\columnwidth]{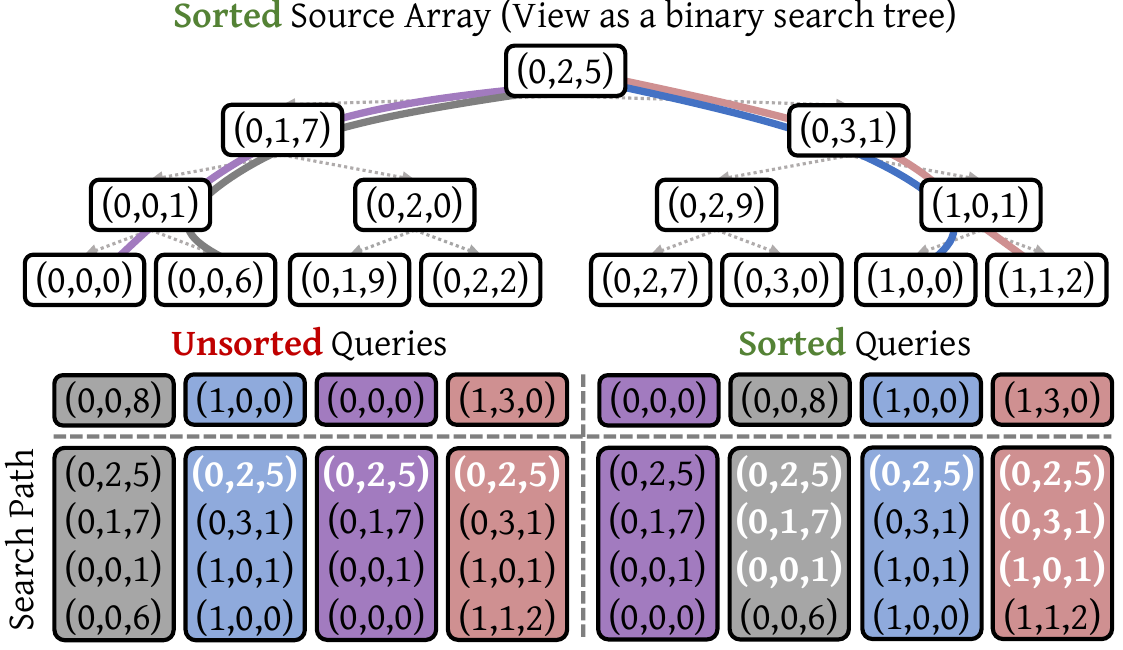}
    \caption{An example binary search execution of four queries, when queries are traversed randomly, i.e., unsorted queries (left) versus via a sorted query approach (right).}
    \label{fig:unsorted-vs-sorted}
\end{figure}

\noindent Figure~\ref{fig:unsorted-vs-sorted} shows an example execution of searching four queries via binary search in the source array (that is visualized as a binary search tree), by traversing queries randomly, i.e., unsorted queries (left), versus via a sorted approach, i.e., sorted queries (right). The annotated white bold coordinates represent the elements of the source array that are common, when executing two adjacent queries. 
In the \emph{unsorted} query execution, when searching for $(1,0,0)$ right after $(0,0,8)$ has been searched, there is \emph{only one} element (i.e., $(0,2,5)$) that is common between the two search paths of adjacent queries.
Instead, in the \emph{sorted} query execution, when searching for $(0,0,8)$ right after $(0,0,0)$ has been searched, there are three common elements (i.e., $(0,2,5)$, $(0,1,7)$ and $(0,0,1)$) in the search paths of adjacent queries. 

The common elements between the search paths of consecutive sorted queries enable two implications for binary search.
First, when executing two consecutive queries with binary search, there is a high probability that the second query accesses the \emph{common} source array elements via the on-chip caches, since common elements might be already cached thanks to executing the first query. 
Thus, the binary search with sorted queries is friendly to GPU memory hierarchy.
Second, each element accessed in the search path corresponds to one comparison between a query and a source array element. Having common elements in search paths means that \emph{the same} source array elements are compared to multiple \emph{sorted} queries (e.g., $(0,2,5)$ is compared to all four sorted queries).
Thus, if we could find the lower bound of the source array element in the query array segment, i.e. the smallest query within the query segment that is no smaller than a source array element, we could avoid the comparisons to that source array element, thus reducing the number of comparisons in binary search scheme with sorted queries.

To this end, \system address two challenges. 
First, binary search with sorted queries necessitates that the source and query arrays need to be sorted, and thus we need to minimize the sorting overheads in both arrays (\textbf{Challenge 1}). 
Second, we need to exploit both the memory friendliness and the aforementioned optimization of reducing comparisons in binary search with sorted queries (\textbf{Challenge 2}).

\subsubsection{Segmented Query Sorting}\label{sec:query-sorting}
\begin{figure}[t]
    \centering
    \includegraphics[width=\columnwidth]{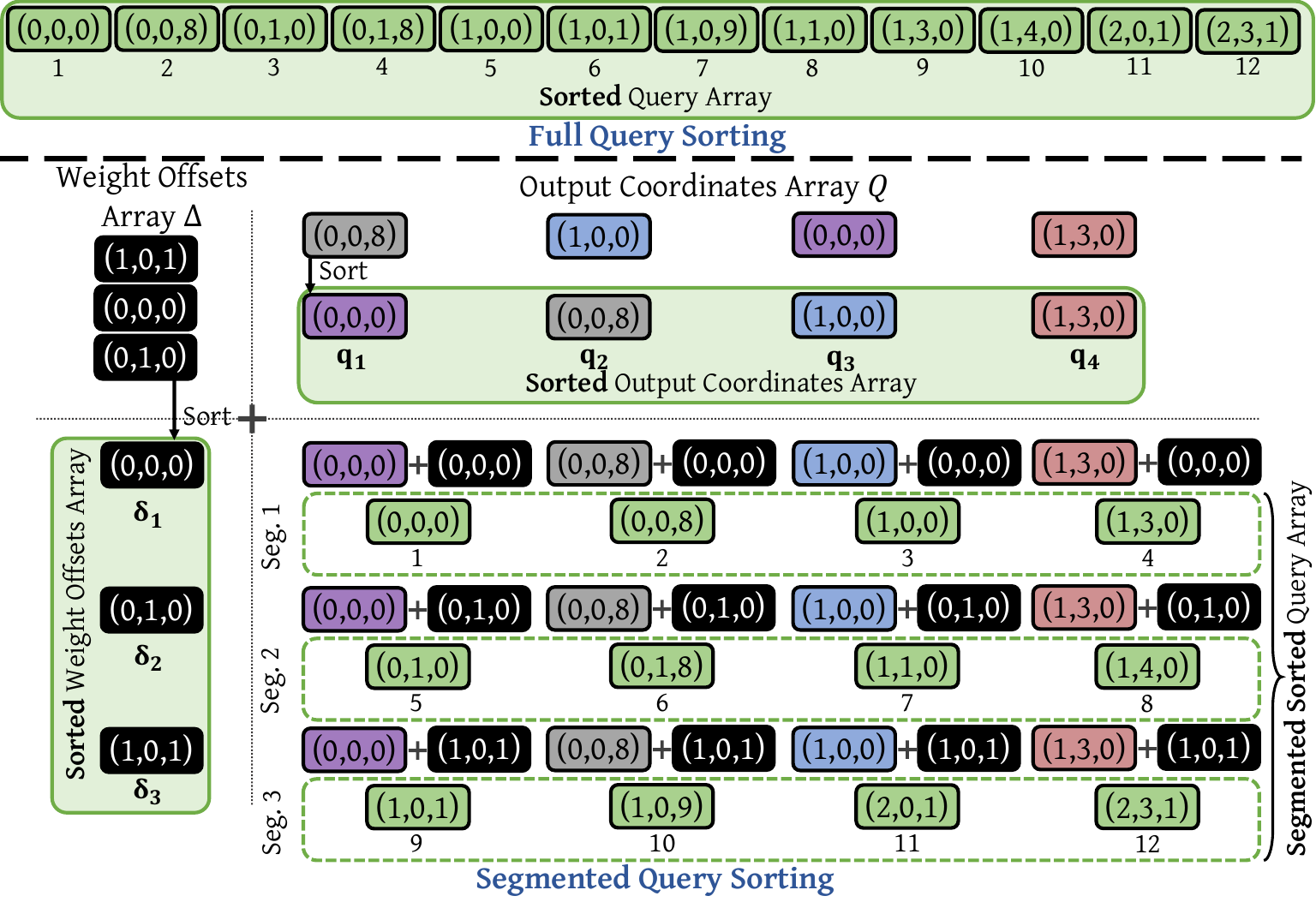}
    \caption{An example of full query sorting (top) and segmented query sorting (bottom). The solid green boxes represent arrays that are materialized in GPU global memory. The dashed green boxes represent arrays that are \emph{not} materialized in memory and their values are calculated on-the-fly.}
    \label{fig:segmented-query-sorting}
\end{figure}

A na\"ive approach to build the kernel map is to materialize all possible queries $\{\mathbf{q_i} + \boldsymbol{\delta}_\mathbf{k}\}$ in a query array, sort the query array and execute binary search for each query within the source array. 
This approach, referred to as \emph{full query sorting}, is depicted in Figure \ref{fig:segmented-query-sorting} top, in which we assume that there are $4$ output coordinates $\mathbf{q_i}$ and $3$ weight offsets $\boldsymbol{\delta}_\mathbf{k}$, thus resulting in $12$ queries, which are perfectly sorted in the full query sorting approach. 
Binary searching with full query sorting is highly cache-friendly, as explained, since searching \emph{sorted} queries in the source array results in many accesses to the same elements of the source array. 
However, full query sorting incurs high sorting overheads: (1) the size of the query array $K^3|\mathcal{Q}|$ is much larger than the size of the source array $|\mathcal{P}|$, and thus sorting the query array causes even higher sorting overheads than that of the source array itself; (2) the large query array needs to be sorted at \emph{each} \SpConv layer of the point cloud network. 
In practice, we found that using full query sorting approach to build the kernel maps of \SpConv layers takes much longer time than using the hash table-based approach of prior \SpConv engines~\cite{MLSYS2022_6512bd43,DBLP:journals/corr/abs-1904-08755}. 
Therefore, we conclude that the full query sorting approach has huge sorting overheads that offset its cache benefits, and this na\"ive approach does not address the Challenge 1.

To minimize sorting overheads in binary search-based kernel map building, we propose \emph{segmented query sorting}, depicted in Figure \ref{fig:segmented-query-sorting} bottom. In the segmented query sorting, we sort the array of the output coordinates  $\mathbf{q_i}$ and the array of the weight offsets $\boldsymbol{\delta}_\mathbf{k}$ \emph{separately}, i.e., we materialize two separate arrays in memory (solid green boxes) and sort each of them, and then we execute all possible queries as \emph{sorted segments} (dashed green boxes): we iterate through all weight offsets, and for each weight offset $\boldsymbol{\delta}_\mathbf{k}$ we on-the-fly create a sorted segment of possible queries (without materializing a new array for the segment) by adding the current weight offset to each sorted output coordinate $\mathbf{q_i}$ of the output coordinate array. For example, in Figure \ref{fig:segmented-query-sorting} the $2$-nd segment is created on-the-fly by adding the $2$-nd sorted weight offset $(0, 1, 0)$ to each sorted output coordinate, i.e., $(0, 0, 0)$, $(0, 0, 8), (1, 0, 0), (1, 3, 0)$, thus the $2$-nd segment comprises of the four elements $(0, 1, 0)$, $(0, 1, 8)$, $(1, 1, 0)$, and $(1, 4, 0)$. 

Segmented query sorting leverages many cache friendly accesses in binary search-based query lookups, while also minimizing the sorting overheads in both source and query arrays, thus addressing \textbf{Challenge 1}. 

On the one hand, since the output coordinate array is sorted, and we create a query segment by adding the same weight offset to each sorted element of the output coordinate array, the produced queries in the query segment are by nature sorted as well (See segments of Figure \ref{fig:segmented-query-sorting}).
In practice, for a typical \SpConv layer with kernel size $K$, the number of weight offsets (i.e., $K^3$) to be sorted for each \SpConv layer is much smaller than the number of output coordinates, i.e., $K^3 \lll |\mathcal{Q}|$.
For example, a typical \SpConv layer has a kernel size $3$, and there are $27$ weight offsets to be sorted, while the number of output coordinates is much larger (e.g., $10^5$).
As a result, there is only a small number of segments, but \emph{a large number of queries} within each segment. 
Thus, segment query sorting has segments with many sorted elements, and enables a sufficiently large number of cache-friendly memory accesses for binary search that are close to that of the full query sorting approach.

On the other hand, segmented query sorting minimizes the sorting overheads for four compelling reasons.

First, weight offsets sorting is \emph{not} in the critical inference path. 
Weight offsets are determined by the \SpConv layer configuration itself (i.e., the kernel size and stride, as discussed in Section \ref{sec:background-definition}) and are independent to the input point cloud data, so they need to be sorted only \emph{once} for each \SpConv layer in the network. 
This sorting is performed as a preprocessing step, when loading the configuration of the \SpConv layer, and has negligible costs, since the number of weight offsets of each \SpConv layer is very small (e.g., $27$ for a typical kernel size $3$).

Second, segmented query sorting sorts the output coordinate array of size $|\mathcal{Q}|$ (e.g., $10^5$), the sorting cost of which is smaller than sorting the whole query array of size $K^3 |\mathcal{Q}|$ (e.g., $27 \times 10^5$) which is needed in the full query sorting approach. 
Moreover, segmented query sorting sorts the output coordinate array only \emph{once} for \emph{all} query segments, which are created on-the-fly (dashed green boxes in Figure \ref{fig:segmented-query-sorting}) and are \emph{not} materialized in memory. Thus, it performs much smaller number of memory accesses for queries compared to the full sorting approach, that first needs to materialize in memory the whole query array before sorting it (solid green box in Figure \ref{fig:segmented-query-sorting} top).

\begin{figure}[t]
    \centering
    \includegraphics[width=\columnwidth]{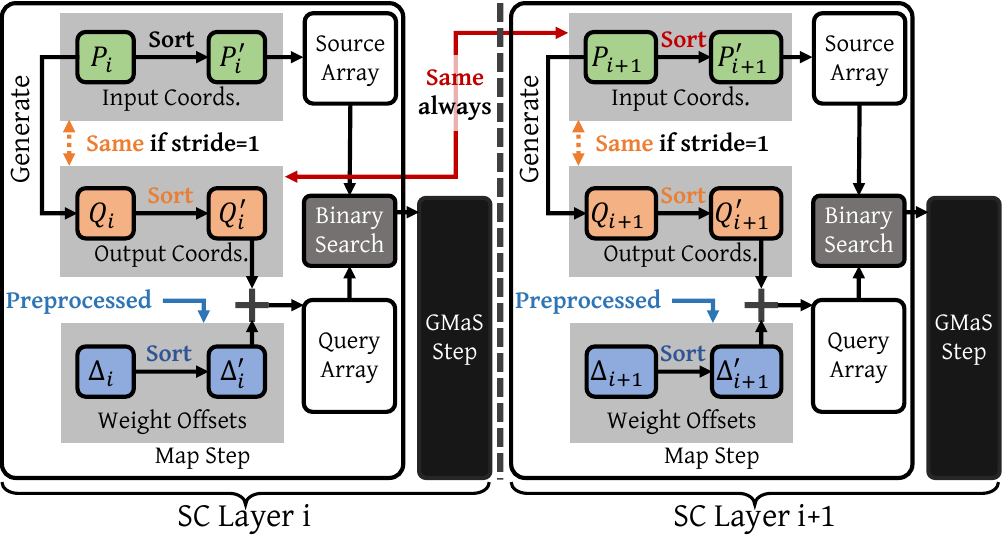}
    \caption{Optimizing sorting overheads of weight offsets and output coordinates in adjacent \SpConv layers.}
    \label{fig:efficient-sorting-coordinates}
\end{figure}

Third, \SpConv models typically have multiple \SpConv layers connected sequentially \cite{DBLP:journals/corr/abs-1904-08755} the one after the other, and the output coordinates of one \SpConv layer are the input coordinates of the subsequent \SpConv layer. Figure \ref{fig:efficient-sorting-coordinates} presents two adjacent \SpConv layers as a part of a large point cloud network.
Segmented query sorting requires the output coordinate array $\mathcal{Q}$ to be sorted in the \Map step of each \SpConv layer to perform binary search-based kernel map building. Thus, by leveraging segmented query sorting, sorting the input coordinate array of the layer $i+1$ is \emph{completely} eliminated, since input coordinate array of the layer $i+1$ is the always same as output coordinate array of the layer $i$, which is already sorted (the red solid arrow in Figure \ref{fig:efficient-sorting-coordinates}). 
This optimization cannot be enabled with the full query sorting approach, since it necessitates a different (separate) array across different \SpConv layers to store and sort the queries, because the weight offsets (or the coordinates) can be different.

Fourth, when the stride of an \SpConv layer is one, the output coordinates $\mathcal{Q}$ are identical to the input coordinates $\mathcal{P}$ (orange dashed arrows in Figure \ref{fig:efficient-sorting-coordinates}), as explained in Section \ref{sec:background-definition}. Therefore, in \SpConv layers with stride 1, we do \emph{not} materialize and sort two separate arrays for the source and query arrays. Instead, we materialize only one array that serves as both sort and query array and sort it only \emph{once}. 
Similarly this optimization cannot be enabled with the full query sorting approach because it necessitates separate query array to store and sort the queries.

Overall, \system significantly minimizes sorting overheads and bulid the kernel map very efficiently  via segmented query sorting.
\system  leverages existing GPU radix sorting libraries~\cite{nvlabsCUBMain} to sort the arrays at low cost. 
In Figure \ref{fig:eval-map-build-time} (Section~\ref{sec:eval-map-step-build}), we show that the building time of \system is faster than that of prior \SpConv engines.

\subsubsection{Double-Traversed Binary Search} 
To solve the Challenge 2, we introduce a novel binary search algorithm that both reduces the comparisons, and efficiently leverages the on-chip memory hierarchy of GPUs when using sorted queries.

\begin{figure}[t]
    \centering
    \includegraphics[width=0.85\columnwidth]{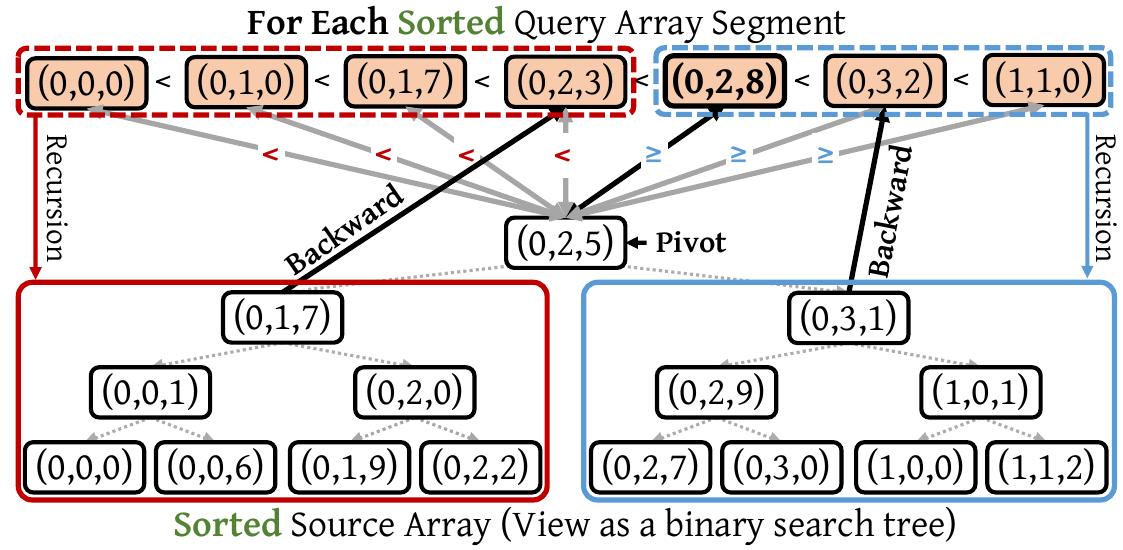}
    \caption{An example of using backward binary search to find the lower bound of the pivot in the query array segment to reduce the number of comparisons, when executing binary search with a sorted query segment in the sorted source array (represented as a binary search tree).}
    \label{fig:redundant-comparison}
\end{figure}

Figure \ref{fig:redundant-comparison} depicts an example of executing binary search with one sorted query array segment into the sorted source array that is represented as a binary search tree. 
Each query in the query array segment needs to be compared with the middle element of the source array, i.e., the element $(0, 2, 5)$, in the first comparison step of the binary search. We refer to such element as the \textbf{pivot}. We observe that as traversing the sorted queries of the query array segment there is \emph{at most one change} from a smaller ``$<$'' element to a larger ``$>$'' element than the pivot, e.g., all queries from $(0, 0, 0)$ to $(0, 2, 3)$ are smaller than the pivot (the red dashed box), and all queries from $(0, 2, 8)$ to $(1, 1, 0)$ are larger than pivot (the blue dashed box). 
Thus, if we find the lower bound of the pivot within the query array segment (i.e., the bold coordinate $(0, 2, 8)$), namely the first element of query array segment that is no smaller than the pivot, we could avoid many comparisons to the pivot for the elements of the query array segment: according to the transitivity property, all queries before the lower bound (red dashed box) will be smaller than the pivot, and all queries after the lower bound (blue dashed box) will be larger than or equal to the pivot. 

\noindent\textbf{Key Idea.}
To find the lower bound of pivot, we apply binary search in a \emph{backward} manner (\textbf{backward binary search}), namely to binary search the pivot in the query segment. 

This key idea can be applied recursively in all elements (pivots) of the source array. 
The sorted source array can be split into the (i) subarray with elements smaller than the pivot (the left subtree with red solid box), and the (ii) subarray with elements no smaller than the pivot (the right subtree with blue solid box). 
The (i) subarray is associated with the left query subarray, i.e., the queries that are smaller than the lower bound (the red dashed box), and the (ii) subarray is associated with the right query array, i.e., the queries that are no smaller than the lower bound (the blue dashed box). 
Then, the backward binary search is applied to the pivot (roots) of the (i) and (ii) source subarrays (subtrees), i.e., $(0, 1, 7)$ and $(0, 3, 1)$, and proceeds recursively to all elements of the source array.
However, recursively applying backward binary search would require many recursive function calls, which limits the degree of execution parallelism and incurs high warp divergence overheads on GPUs. 

\begin{figure}[t]
    \centering
    \includegraphics[width=\linewidth]{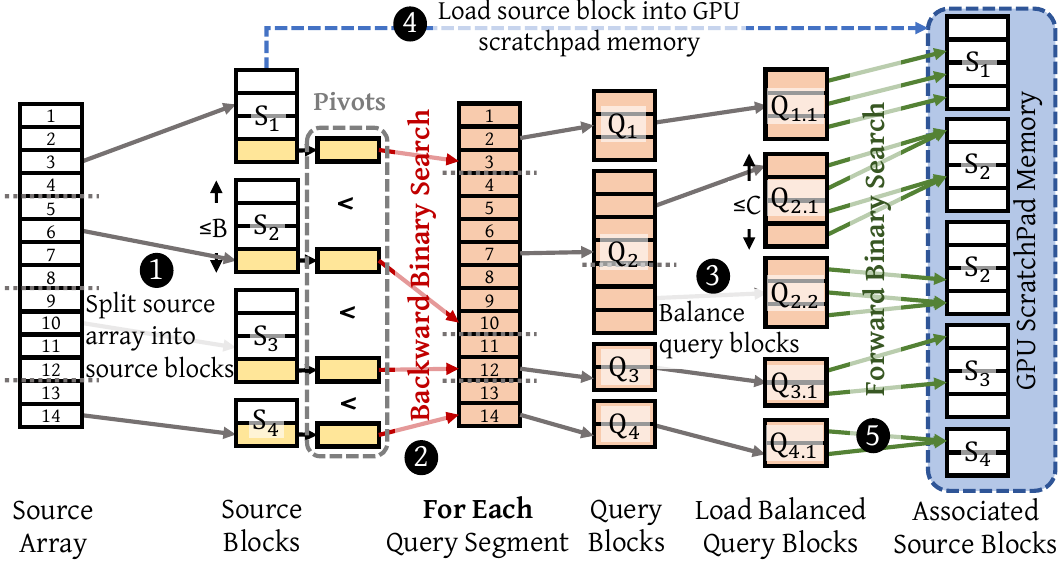}
    \caption{Double-traversed binary search execution steps.}
    \label{fig:double-traversed-binary-search}
\end{figure}

To this end, we consider only one level of backward binary search and propose \emph{double-traversed binary search} algorithm for kernel map building, that consists of two steps. Figure \ref{fig:double-traversed-binary-search} presents the execution steps of our proposed algorithm.

\noindent\textbf{Backward Binary Search.}
Instead of using only one pivot in source array, we select multiple pivots, and split the source and query arrays to multiple blocks.
We first partition the source array into multiple \emph{source blocks} \circled{1}, each of them has a size that is no larger than a hyperparameter $B$. Then, we use the last elements of each source block as \emph{pivots} to split the query segment into multiple \emph{query blocks}: for each pivot of a source block, we perform backward binary search to the sorted query segment \circled{2} to find the subset of consecutive queries, i.e., the query block, that is associated with that source block. Thus, the query segment is split in multiple query blocks, the number of which is equal to the number of source blocks in the source array.

\noindent\textbf{Forward Binary Search.}
We observe that the query blocks are data dependent, thus their sizes might significantly vary across them. To enable load balance across GPU threads, we balance query blocks \circled{3} by further splitting all query blocks that have size larger than than a hyperparameter $C$ (e.g., $Q_2$). This way all query blocks are load balanced, i.e., having size that is no larger than $C$. 
Then, we assign one CUDA thread block to each query block, where the thread block processes all queries of that query block by performing forward binary search to the associated source block. 
To do so, each CUDA thread block first loads the associated source block into the GPU scratchpad memory \circled{4} to minimize the number of global memory accesses. Then, each thread of the CUDA thread block executes forward binary search \circled{5} to check the existence of each query in the query block within the source block.

\system achieves very low memory access costs, while also reduces the number of comparisons (\textbf{Challenge 2}).
First, both backward and forward binary search provide high memory efficiency. The backward binary search is highly cache friendly, since the pivots of source blocks are \emph{sorted}, thus it is treated as binary search with sorted queries (Section \ref{sec:opt-map-step}). 
The forward binary search also provides high memory benefits, since we only access global memory, when fetching the source block to scratchpad memory and the query block to register files, while accessing the elements within both the source and the query blocks has a sequential memory access pattern.
Second, we reduce the search range for each query block from the whole source array of size $|\mathcal{Q}|$ to the source block of size $B$ by the backward binary search.
This way, we significantly reduce the number of comparisons performed by our algorithm.

\subsubsection{Computational Complexity of Segmented Sorting Double-Traversed Binary Search}
\label{sec:bs-complexity}
We use Concurrent-Read-Exclusive-Write (CREW) as the Parallel RAM model for computational complexity analysis (i.e., work and time complexity).
For simplicity, we assume all source blocks and  load balanced query blocks have the same sizes of $B$ and $C$, respectively.
Under this assumption, we provide only the work complexity analysis, since both the backward and the forward binary search can be straightforwardly parallelized, and the time complexity is simply the work complexity divided by the number of processors.

Let $K$, $|\mathcal{P}|$, and $|\mathcal{Q}|$ be the kernel size of the \SpConv layer, the number of input and output coordinates, respectively.
In the backward binary search, for each of the $\left\lceil \frac{|\mathcal{P}|}{B}\right\rceil$ source block, loading the last element of the source block takes $\mathcal{O}(1)$ time and searching it in one sorted query segment array takes $\mathcal{O}\left(\log|\mathcal{Q}|\right)$ comparisons.
In the forward binary search, for each of the $\mathcal{O}\left(\frac{|\mathcal{P}|}{B} + \frac{|\mathcal{Q}|}{C}\right)$ load balanced query block\footnote{Let $x_i$ denote the size of the unbalanced query block for the $i$-th source block, we have $\sum_{i=1}^{\left\lceil \frac{|\mathcal{P}|}{B}\right\rceil} \left\lceil \frac{x_i}{C}\right\rceil \in \mathcal{O}\left(\frac{|\mathcal{P}|}{B} + \frac{|\mathcal{Q}|}{C}\right)$ load balanced query blocks.}, loading the source and query block from global memory takes $\mathcal{O}\left(C + B\right)$ time, and the in-scratchpad binary search takes $\mathcal{O}\left(C\log B\right)$ time.
In \SpConv, $|\mathcal{P}|$ has the same order of magnitude with $|\mathcal{Q}|$. By configuring $B = \frac{|\mathcal{P}|}{|\mathcal{Q}|} \log |\mathcal{Q}|$ and $C = \sqrt{\frac{|\mathcal{Q}|}{|\mathcal{P}|\log B}}B$, the work complexity of the segmented sorting double-traversed binary search is:
\begin{equation}\label{eq:bs-complexity}
    \begin{split}
        &\textstyle\mathcal{O}\left(K^3\left(\frac{|\mathcal{P}|}{B}\log|\mathcal{Q}| + \left(\frac{|\mathcal{P}|}{B} + \frac{|\mathcal{Q}|}{C}\right)\left(B+C\log B\right) \right)\right)\\
        = &\textstyle \mathcal{O}\left(K^3|\mathcal{Q}|\log \log |\mathcal{Q}|\right)
    \end{split}
\end{equation}
\system can thus achieve a computational complexity close to that of hash table-based kernel map building, i.e., $\mathcal{O}(K^3|\mathcal{Q}|)$.

\subsubsection{\system's Selection of Hyperparameters $B$ and $C$} \system carefully chooses the hyperparameter $B$ and $C$.
Intuitively, hyperparameter $B$ balances the trade-off between the execution times of the forward and the backward binary search: a larger value of $B$ results in fewer but larger source blocks, which decreases the number of comparisons in the backward binary search, but increases the number of comparisons in the forward binary search, and vice versa.
Hyperparameter $C$ balances the trade-off between data movement and the load balance in the forward binary search: a larger value of $C$ results in fewer but larger query blocks, which decreases the data movement for copying the associated source block to the scratchpad memory, but increases the load imbalance among CUDA thread blocks, and vice versa.
In Figure \ref{fig:eval-map-hyperparameter}, we provide a sensitivity study on the values of the hyperparameters $B$ and $C$ using various GPUs, and find that with thread block size of $128$, configuring $B = 256$ and $C = 512$ (default \system's values) consistently achieves the best performance among all evaluated GPUs and datasets. 
For flexibility, we expose $B$ and $C$ as configurable hyperparameters to users.

\subsection{Optimizing the \GMulS Step}\label{sec:opt-gmas-step}

In this section, we describe the optimizations and trade-offs of \system in the \GMulS step.

\subsubsection{Autotuned Gather/Scatter} 
We summarize \system's algorithm of the Gather operation in Algorithm~\ref{alg:gather}. The Scatter operation can be conducted similarly to Gather. 

\begin{algorithm}[t]\small
\captionsetup{font=small}
\caption{The Gather operation}\label{alg:gather}
\begin{algorithmic}[1]
\Require Weight offsets $\boldsymbol{\Delta}$, Input channel size $C_\text{in}$, Input coordinates $\mathcal{P} = \{\mathbf{p_i}\}$, Input buffer array $\{\mathbf{b_i}\}$, Input metadata table $IMT$, Gather tile size $T$
\Ensure Input feature vectors $\{\mathbf{F^\mathcal{P}_i}\}$
\For{$t \gets 0, 1, \dots, \frac{C_\text{in}}{T} - 1$ \textbf{in parallel}}\label{alg:for_outer}
    \For{$\mathbf{p_i} \in \mathcal{P}$ \textbf{in parallel}}
        \State Read from the $t$-th tile of $\mathbf{F^{\mathcal{P}}_i}$ to $\mathbf{v}$ (in register files)
        \For{$\boldsymbol{\delta}_\mathbf{k} \in \boldsymbol{\Delta}$}\label{alg:for_inner}
            \State $\text{index} \gets \mathrm{GetInputBufferIndex}(IMT, {\color{blue}\boldsymbol{\delta}_\mathbf{k}, \mathbf{p_i}})$\label{alg:gather-invariant}
            \If{$\text{index} \neq \varnothing$}
                \State Write $\mathbf{v}$ to the $t$-th tile of $\mathbf{b_{index}}$
            \EndIf
        \EndFor
    \EndFor
\EndFor
\end{algorithmic}
\end{algorithm}

With a given tile size $T$, the Gather operation assign one CUDA thread to each feature channel tile and each input coordinate, which achieves a parallelism of $\frac{C_\text{in}}{T} \times |\mathcal{P}|$.
As shown in blue at line \ref{alg:gather-invariant}, we observe that the accesses in input metadata table are not related to the tile index $t$, which implies the all the $\frac{C_\text{in}}{T}$ accesses are to the same entry in the metadata table within the same tile.
Hence, on the one hand, increasing the tile size reduces indexing costs, namely the number of accesses to the metadata table, i.e., $\frac{C_\text{in}}{T} \times |\mathcal{P}| \times K^3$.
However, on the other hand, increasing the tile size $T$ also reduces the execution parallelism $\frac{C_\text{in}}{T} \times |\mathcal{P}|$. 
As a result, we might not saturate the GPU, especially when the number of input/output coordinates is small or when we use powerful GPUs with a large number of processing units. 

To trade-off between indexing costs and execution parallelism, we propose to autotune the tile size for each Gather and Scatter operation of the model.
In Algorithm \ref{alg:autotune}, we demonstrate how \system autotunes the Gather operation (the Scatter operation is autotuned similarly).
Specifically, we sample a few point clouds from the dataset and feed them to the \SpConv network (line \ref{alg:line_sample_pc}). Then, for each \SpConv layer in the network, we create the metadata tables for these few point clouds and use them to find the best-performing tile size (line \ref{alg:line_metadata}).
Next, we exhaustively search all possible tile sizes (line \ref{alg:line_search}), i.e., the divisors of $C_\text{in}$, 
and select the tile size with the minimum latency (line \ref{alg:line_update_gather}).
Note that this autotuning process only happens once, before running the inference (pre-processing cost), and does not introduce significant overhead (less than 2 minutes) as presented in Section \ref{sec:eval-setup}. 
\begin{algorithm}[t]\small
\captionsetup{font=small}
\caption{Autotuning the Gather operations.}\label{alg:autotune}
\begin{algorithmic}[1]
\Require A \SpConv network $\mathcal{M}$ of $n$ layers $\mathcal{L}_i$, A point cloud dataset for tuning $\mathcal{D}$, The rounds of tuning $R$
\Ensure The tuned \SpConv network $\mathcal{M}^\text{tunned} = \{\mathcal{L}^\text{tunned}_i\}$
\State $\mathcal{D}_\text{Sampled} \gets $Sample a few point clouds from the dataset $\mathcal{D}$\label{alg:line_sample_pc}
\For {each layer $\mathcal{L}_i$}
    \State Compute input metadata tables $IMT^{(i)}$ based on $\mathcal{D}_\text{Sampled}$\label{alg:line_metadata}
    \State $T^*_\text{Gather} \gets \varnothing$
    \For{each divisor $T$ of $\mathcal{L}_i$'s input channel size $C_\text{in}$}\label{alg:line_search}
        \State Profile \textsc{Gather} for $R$ rounds with $T$ and $IMT^{(i)}$
        \State Update $T^*_\text{Gather}$ with $T$ if the latency is smaller\label{alg:line_update_gather}
    \EndFor
    \State $\mathcal{L}^\text{tunned}_i \gets \mathcal{L}_i$ with $T^*_\text{Gather}$ in the Gather operation
\EndFor
\end{algorithmic}
\end{algorithm}

\subsubsection{Padding-Efficient GEMM Grouping}
To improve hardware utilization in transforming input features to output features, prior \SpConv engines~\cite{MLSYS2022_6512bd43} use zero-padding in input and output features to achieve better compute regularity and low launch overheads in GEMM operations.
However, we observe that the padding strategy proposed in prior works is still inefficient.
To tackle this, we propose to reorder the weights based on the size of their corresponding GEMM operations, i.e., in non-decreasing order of the number of input features to be multiplied by each weight.
After reordering, we employ a similar adaptive policy for grouping adjacent GEMM operations, as proposed by prior works~\cite{MLSYS2022_6512bd43}. 

Intuitively, after we reorder the weights, adjacent weights will have the same or very similar sizes in GEMM operations.
Thus, there is only a small amount of padding with zero values needed to have the same sizes/heights in the GEMM's operands, which consequently reduces unnecessary data accesses and computations.
Note that the reordering requires sorting the GEMM sizes and permuting the weights. 
However, we found this sorting incurs negligible overhead, being less than $4\%$ of the layer execution time. 
This sorting overhead is accounted for in our evaluations, and in Section~\ref{sec:eval-layerwise} we show that \system has better layerwise performance than prior works. 
To further improve hardware utilization, we parallelize all GEMM kernels by executing them on a pool of CUDA streams~\cite{nvidiaCUDAStream}.
We set the stream pool size $s$ to $4$ in \system, since we found that increasing $s$ larger than 4 results in no further performance speedups.
\section{Evaluation}\label{sec:eval}

\subsection{Methodology}\label{sec:eval-setup}

We followed existing common practice  \cite{MLSYS2022_6512bd43,10.1145/3582016.3582047,won2023unified,DBLP:journals/corr/abs-2201-05072,10.1145/3575693.3575702} to develop methodology to evaluate \system's SC executions.

\begin{figure*}
    \centering
    \resizebox{\linewidth}{!}{\input{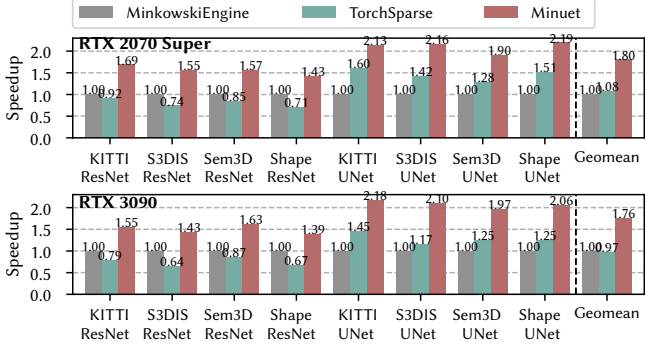}}
    \caption{End-to-end performance of all \SpConv engines using various networks, real datasets and GPU architectures.}
    \label{fig:eval-end-to-end}
\end{figure*}

\noindent\textbf{Platforms.} We evaluate \system on 4 NVIDIA GPU servers, RTX 2070 Super (8 GB), RTX 2080 Ti (11 GB), RTX 3090 (24 GB), and Tesla A100 (80 GB).
All GPU servers have CUDA 11.8.0 and PyTorch 2.0.0 installed.
Unless otherwise noted, we present detailed evaluation results on \MinuetMainGPU.

\noindent\textbf{Baselines.} We compare \system with two state-of-the-art \SpConv engines: 
(1) MinkowskiEngine~\cite{MLSYS2022_6512bd43},
and (2) TorchSparse~\cite{DBLP:journals/corr/abs-1904-08755}.
In \system, we account for both the overheads of sorting coordinates and GEMM reordering in end-to-end, layerwise, and \GMulS step evaluations. 
However, for TorchSparse and \system, we exclude the autotuning time as both autotuning processes happen only \emph{once} and are before the inference. 
The autotuning process of \system takes less than 2 minutes to finish on all datasets evaluated.

\noindent\textbf{Neural Networks.} We evaluate two representative and commonly used 3D point cloud neural networks:
(1) SparseResNet21 (\emph{ResNet})~\cite{3DSemanticSegmentationWithSubmanifoldSparseConvNet,SubmanifoldSparseConvNet} that serves as the backbone for the widely used CenterPoint 3D object detector \cite{Yin_2021_CVPR}; 
and (2) MinkUNet42 (\emph{UNet})~\cite{DBLP:journals/corr/abs-1904-08755} that achieves top-level accuracy in processing 3D point cloud data.

\noindent\textbf{Datasets.} We evaluate four large-scale point cloud datasets: (1) SemanticKITTI Dataset (\emph{KITTI})~\cite{DBLP:journals/corr/abs-1904-01416} which includes outdoor LiDAR scans for self-driving scenarios, 
(2) Stanford 3D Indoor Scene Dataset (\emph{S3DIS})~\cite{Armeni_2016_CVPR} which labels 3D objects in indoor areas, 
(3) Semantic3D Dataset (\emph{Sem3D})~\cite{hackel2017isprs} which is a large-scale dataset for outdoor scenes,
and (4) ShapeNetSem Dataset (\emph{Shape})~\cite{shapenet2015,savva2015semgeo} which contains large-scale point clouds for 3D models.
Note that to feed a point cloud to \SpConv networks, the floating-point number coordinates are first voxelized~\cite{DBLP:journals/corr/abs-1904-08755} into integers. 
After voxelization, the average sparsity\footnote{This is defined as the number of non-zero data points divided by the bounding volume and averaged over all point clouds in the dataset.} is $0.04\%$, $2\%$, $0.03\%$, and $10\%$ for the \emph{KITTI}, \emph{S3DIS}, \emph{Sem3D}, and \emph{Shape} datasets, respectively.
To study how the \system's optimizations in the \Map step are affected by data distribution and sparsity patterns (Figure \ref{fig:eval-sparsity}, \ref{fig:eval-map-query-time}, and \ref{fig:eval-map-build-time}), we generate a random artificial dataset: we vary the voxelization process in \emph{Sem3D} dataset to provide different numbers of non-zero points in each point cloud in the dataset.

\subsection{End-to-End Performance}

\noindent\textbf{Total Speedup.}
Figure~\ref{fig:eval-end-to-end} compares the end-to-end speedup of all \SpConv engines, when executing the neural networks on various datasets and GPUs. 
We make two key observations.
First, there is no clear winner between MinkowskiEngine and TorchSparse:  MinkowskiEngine  outperforms TorchSparse on  \emph{ResNet} network, while it performs worse than TorchSparse on \emph{UNet} network.
This is because MinkowskiEngine is specialized for small channel size \SpConv layers, which dominate the \emph{ResNet} network.
Second, \system consistently outperforms prior \SpConv engines, by \Eval{MinuetE2EVSMinkowskiEngineAvg} on average (up to \Eval{MinuetE2EVSMinkowskiEngineMax}) over MinkowskiEngine, and \Eval{MinuetE2EVSTorchSparseAvg} on average (up to \Eval{MinuetE2EVSTorchSparseMax}) over TorchSparse, for all neural networks, datasets, and GPU systems.
Noticeably, \system achieves close to $2\times$ speedup on \emph{UNet} on RTX 2070, RTX 2080 Ti, and RTX 3090 over MinkowskiEngine thanks to low-cost data accesses in the \Map step and high parallelism and hardware utilization in the \GMulS step. 
Overall, we conclude that \system achieves the best performance over prior state-of-the-art \SpConv engines across various sparse point cloud networks, real datasets, and even when using different GPU architectures.

\noindent\textbf{Sensitivity on Point Cloud Density.}
\begin{figure}[t]
    \centering
    \resizebox{\linewidth}{!}{\input{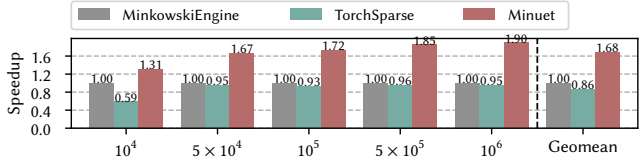}}
    \caption{End-to-end performance of all \SpConv engines with varying point cloud density.}
    \label{fig:eval-sparsity}
\end{figure}
We evaluate \system on a random synthetic dataset to understand how \system generalizes to point clouds with different input densities.
Specifically, we randomly generate point clouds within a fixed bounding volume ($400 \times 400 \times 400$) and vary the number of non-zero points from $10^4$ to $10^6$ to achieve different input densities. 
Figure~\ref{fig:eval-sparsity} shows the end-to-end speedup with MinkowskiEngine and TorchSparse, where \system consistently outperforms existing \SpConv engines on various input densities by on average \Eval{MinuetSparsityAvg} (up to \Eval{MinuetSparsityMax}).

\begin{figure}[t]
    \centering
    \resizebox{\linewidth}{!}{\input{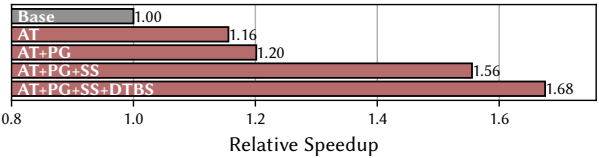}}
    \caption{Performance breakdown of the key ideas in \system, where AT stands for Autotuned Gather/Scatter, PG for Padding-Efficient GEMM Grouping, SS for Segmented Query Sorting, and DTBS for Double-Traversed Binary Search.}
    \label{fig:eval-breakdown}
\end{figure}

\noindent\textbf{Speedup Breakdown.}
To understand how each key idea of \system contributes to the final performance, we evaluate performance by incrementally enabling the four key ideas proposed in \system in Figure~\ref{fig:eval-breakdown}.
We draw two conclusions. 
First, all \system's four key ideas significantly contribute to the final end-to-end performance. 
Second, the most significant speedup comes from the segmented query sorting, which shows the superiority of using the sorted key-value array instead of hash tables in \SpConv execution.

\subsection{Layerwise Performance}\label{sec:eval-layerwise}

\begin{figure}[t]
    \centering
    \resizebox{\linewidth}{!}{\input{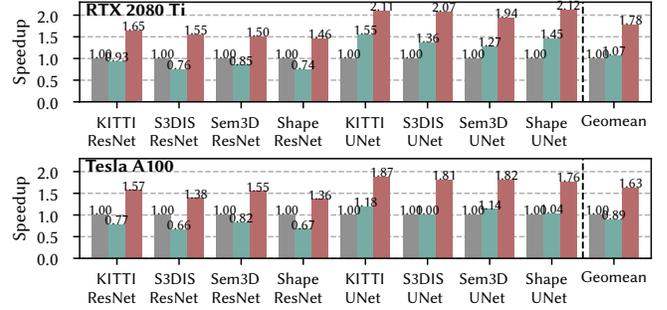}}
    \caption{Layerwise speedup of \SpConv engines averaged across all real datasets when varying the input and output channel sizes in the SC layers.}
    \label{fig:eval-layerwise}
\end{figure}

Figure~\ref{fig:eval-layerwise} compares the layerwise performance of all \SpConv engines on the most commonly used \SpConv layer configurations.
The $x$-axis corresponds to an \SpConv layer with $C_\text{in}$ input and $C_\text{out}$ output channels. 
We calculate the geometric mean across all real datasets for each \SpConv layer, and the last group column shows the geometric mean averaged across all layers.

We draw two findings.
First, TorchSparse performs worse than MinkowskiEngine on \SpConv layers with small channel sizes (e.g., $(4, 16)$), while it performs better on layers with larger channel sizes (e.g., $(128, 128)$). 
This is due to a specialized dataflow that is optimized for small channel sizes in MinkowskiEngine~\cite{MLSYS2022_6512bd43}.
Second, \system significantly outperforms the MinkowskiEngine by on average \Eval{MinuetLayerwiseVSMinkowskiEngineAvg} speedup (up to \Eval{MinuetLayerwiseVSMinkowskiEngineMax}) and TorchSparse by on average \Eval{MinuetLayerwiseVSTorchSparseAvg} speedup (up to \Eval{MinuetLayerwiseVSTorchSparseMax}) in all layer configurations. 
Thus, \system achieves the best performance in various configurations of \SpConv layers. 

\subsection{Performance of the \Map Step}\label{sec:eval-map-step}

\begin{figure}[t]
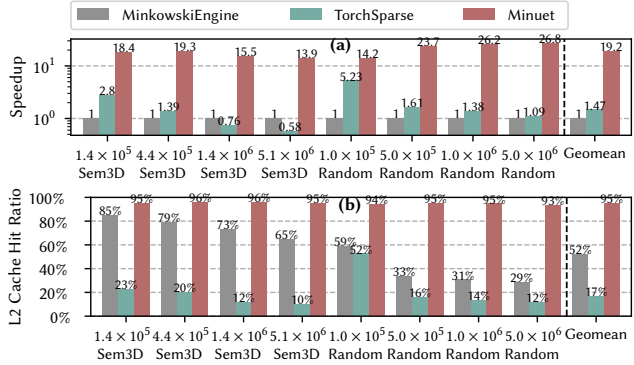

    \centering
    \begin{subfigure}[b]{\linewidth}
        \centering
        \resizebox{\linewidth}{!}{\input{figures/figure3-mapping-query-time.\MinuetMainGPUCode.pgf}}
    \end{subfigure}\\
    \begin{subfigure}[b]{\linewidth}
        \centering
        \resizebox{\linewidth}{!}{\input{figures/figure4-mapping-hit-ratio.\MinuetMainGPUCode.pgf}}
    \end{subfigure}%
    \caption{(a) Normalized speedup and (b) L2 cache hit ratio of the query process to build the kernel map in \SpConv when varying the dataset and number of points in the point cloud.}
    \label{fig:eval-map-query-time}
\end{figure}

\noindent\textbf{Query Process.}
Figure~\ref{fig:eval-map-query-time} compares (a) the execution time and (b) the L2 cache hit ratio achieved (collected with NVIDIA Nsight Compute \cite{NsightCompute}) by \SpConv engines in the query process of the \Map step using the \emph{Sem3D} dataset and an artificial randomly generated dataset (\emph{Random}), that has similar sparsity and number of points with \emph{Sem3D}.
Note that \system's execution time includes the total binary search algorithm proposed, while the L2 cache hit ratio presented represents only the dominating forward binary search process (more than $90\%$ in the total time of the \Map step).
In the $x$-axis, we present the number of points in the input point cloud and the dataset to which the input point cloud belongs.

We make two key observations.
First, thanks to our novel highly memory-efficient binary search algorithm, \system achieves a very high L2 hit ratio, i.e., more than 95\% in all datasets, and provides superior performance benefits in the \Map step: it has \Eval{MinuetMapVSMinkowskiEngineAvg} speedup on average (up to \Eval{MinuetMapVSMinkowskiEngineMax}) and \Eval{MinuetMapVSTorchSparseAvg} speedup on average (up to \Eval{MinuetMapVSTorchSparseMax}) over the hash table implementations of MinkowskiEngine and TorchSparse, respectively.
Second, we observe that as the number of points increases, the cache hit ratio of hash table-based implementations decreases significantly. 
This is because as the number of points increases, the hash table requires larger memory footprint to access the stored input coordinates, which are less likely to remain in the on-chip caches during the query execution. In contrast, \system's segmented sorting double-traversed binary search significantly outperforms hash table-based solutions of prior \SpConv engines, and provides a robust solution, since its performance benefits remain across various numbers of inputs points.

\begin{figure}[t]
    \centering
    \resizebox{\linewidth}{!}{\input{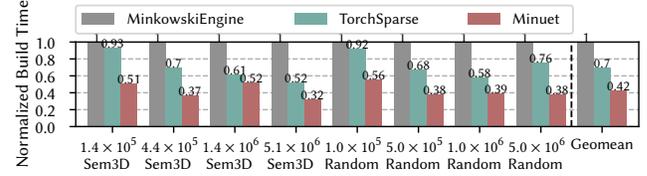}}
    \caption{Building time overhead of the source array when varying the real dataset and the number of points in the input point cloud.}
    \label{fig:eval-map-build-time}
\end{figure}

\noindent\textbf{Build Process.}\label{sec:eval-map-step-build}
Figure~\ref{fig:eval-map-build-time} compares the time for the build process of the \Map step, i.e., the time to build hash tables in MinkowskiEngine and TorchSparse engines, and the time to sort the input/output coordinates in \system. 
By leveraging the high-performance CUDA radix sorting libraries (e.g., NVIDIA CUB~\cite {nvlabsCUBMain}), \system achieves lower building time overhead compared to prior \SpConv engines, and thus sorting cost for the coordinates in our proposed segmented sorting double traversed binary search has negligible overhead.

\begin{figure}[t]
    \centering
    \resizebox{\linewidth}{!}{\input{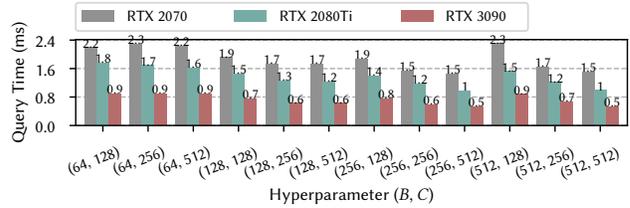}}
    \caption{Query time in the source array, when varying the values of \system's $B$ and $C$ parameters.}
    \label{fig:eval-map-hyperparameter}
\end{figure}

\noindent\textbf{\system's Hyperparameters.}\label{sec:eval-map-step-hyperparameter}
Figure~\ref{fig:eval-map-hyperparameter} shows query time for building kernel maps in the \Map step, when varying the values of the \system's $B$ and $C$ parameters (Section~\ref{sec:opt-map-step}) on three different GPU architectures.
We observe that the best-performing $B$ and $C$ values are not significantly affected by the GPU architecture characteristics, and we choose $B = 256$ and $C = 512$ 
as \system's default values, since they always provide the best performance on all GPU architectures.

\subsection{Performance of the \GMulS Step}

\begin{figure}[t]
    \centering
    \resizebox{\linewidth}{!}{\input{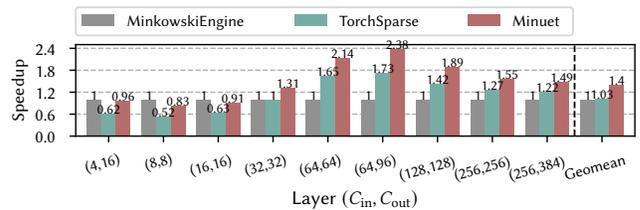}}
    \caption{Normalized speedup in the \GMulS step averaged over all real datasets when varying the input and output channel sizes in \SpConv layers.}
    \label{fig:eval-gather-gemm-scatter}
\end{figure}
\noindent\textbf{\GMulS Execution Time.} 
Figure~\ref{fig:eval-gather-gemm-scatter} compares performance of all \SpConv engines in the \GMulS step in different \SpConv layer configurations, i.e., varying the number of input and output channels.
First, \system on average outperforms prior \SpConv engines. Across all different layer configurations, \system achieves on average \Eval{MinuetGMulSVSMinkowskiEngineAvg} speedup (up to \Eval{MinuetGMulSVSMinkowskiEngineMax}) and  \Eval{MinuetGMulSVSTorchSparseAvg} (up to \Eval{MinuetGMulSVSTorchSparseMax}) over  MinkowskiEngine and TorchSparse, respectively.
This is because \system tunes the tile size on-the-fly and reduces the padding overheads in GEMM operations.
Our evaluations show that TorchSparse incurs on average \Eval{TorchSparseGEMMPadding} padding overhead and launches on average \Eval{TorchSparseGEMMCount} GEMM kernels, while \system has \Eval{MinuetGEMMPadding} padding overhead and launches \Eval{MinuetGEMMCount} GEMM kernels.
Second, we observe that \system's \GMulS step performs slightly worse than MinkowskiEngine (up to $17\%$ slowdown) due to its dedicated optimizations for small channels~\cite{MLSYS2022_6512bd43}.
Overall, we conclude that \system's optimizations in \GMulS step effectively reduce unnecessary data accesses and computations.

\begin{figure}[h]
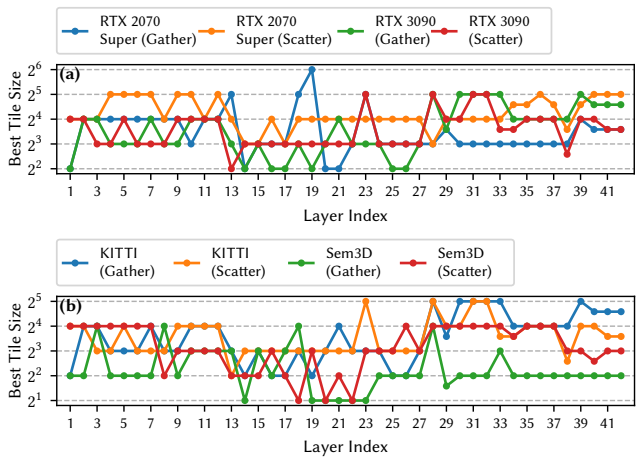

    \centering
    \resizebox{\linewidth}{!}{\input{figures/figure8b_tile_size-diff-gpu.pgf}}
    \resizebox{\linewidth}{!}{\input{figures/figure8c_tile_size-diff-dataset.\MinuetMainGPUCode.pgf}}
    \caption{Best-performing tile size in Gather and Scatter operations of each of the $42$ layers in the MinkUNet42 network \cite{DBLP:journals/corr/abs-1904-08755} on different (a) GPU architectures and (b) datasets.}
    \label{fig:eval-gather-scatter-hyperparameter-diff-gpu}
\end{figure}

\noindent\textbf{\system's Best Performing Tile Size.}
Figure~\ref{fig:eval-gather-scatter-hyperparameter-diff-gpu} presents the best-performing tile size in Gather and Scatter operations of different \SpConv layers of the MinkUNet42 \cite{DBLP:journals/corr/abs-1904-08755} network on various GPUs and datasets, respectively.
We draw two findings.
First, we observe that the best-performing tile size significantly varies across different  datasets and  GPU architectures. This key finding justifies the necessity to tune this parameter in  Gather and Scatter for each dataset and GPU architecture. 
Second, we find that the best-performing tile size also varies across different layers of the network, which indicates that the tile size needs to be re-configured for each \SpConv layer \emph{separately}. 
In contrast, prior works use a fixed tile size (i.e., $4$) in all cases (i.e., \SpConv layers, input dataset and GPU architecture), and thus they are still quite inefficient compared to \system. We conclude that \system's autotuning strategy for tile size enables high system efficiency on various datasets and GPU architectures, and provides a versatile solution to variable layer characteristics of different point cloud networks.
\section{Related Work}
\system is the first work that accelerates \SpConv execution on GPUs by (i) proposing a memory-efficient kernel map building, that highly utilizes the on-chip memory hierarchy of GPUs, and (ii) reducing redundant data accesses in \GMulS step via a batched scheme for the metadata indexing of the input/output feature vectors and a sorted grouping of GEMM operations. We briefly discuss prior work.

\subsubsection*{\SpConv Engines.} Only a few prior works~\cite{DBLP:journals/corr/abs-1904-08755,MLSYS2022_6512bd43,spconv2022} improve the performance of \SpConv execution. MinkowskiEngine~\cite{DBLP:journals/corr/abs-1904-08755} is the first work that proposes a generalized \SpConv to process point clouds and provides an open-sourced \SpConv library. SpConv~\cite{spconv2022} improves the performance of \SpConv by leveraging data locality in GEMM operations. TorchSparse~\cite{MLSYS2022_6512bd43} is the latest optimized \SpConv engine that achieves high system performance by padding and grouping the GEMM operations to improve computation regularity. Our evaluations demonstrate that \system significantly outperforms the prior state-of-the-art \SpConv engines~\cite{MLSYS2022_6512bd43,DBLP:journals/corr/abs-1904-08755} by effectively reducing expensive memory accesses in \Map step and redundant data accesses in \GMulS step. 
\system is also the first work that optimizes the \Map step in \SpConv by highly utilizing the on-chip memory hierarchy on GPUs. 
Finally, PointAcc~\cite{DBLP:journals/corr/abs-2110-07600} proposes a hardware accelerator for point cloud analytics, while \system is a software-based \SpConv engine tailored to modern GPU architectures. 

Concurrent to the submission of this work, PCEngine \cite{hong2023exploiting} and TorchSparse++ \cite{Tang_2023_CVPR} propose to adaptively select dataflows \cite{spconv2022,DBLP:journals/corr/abs-1904-08755,MLSYS2022_6512bd43} for \SpConv execution in the \GMulS step. 
\system is orthogonal to these works.
(1) In \Map step, PCEngine and TorchSparse++ rely on hash tables for building kernel maps, and thus still suffer from expensive data accesses (Shortcoming \#1). 
(2) In \GMulS step, PCEngine and TorchSparse++ still use a fixed tile size in Gather/Scatter operators, thus they suffer from either high indexing costs or limited execution parallelism (Shortcoming \#2), when the Gather-GEMM-Scatter dataflow is selected for \SpConv execution.
PCEngine compresses the kernel map tables \cite{hong2023exploiting} and in turn reduces redundant iterations on weight offsets (line \ref{alg:for_inner} in Algorithm \ref{alg:gather}), which is orthogonal to \system, since \system reduces redundant iterations on feature tiles (line \ref{alg:for_outer} in Algorithm \ref{alg:gather}).
Thus, we conclude that \system's proposed segmented sorted double-traversed binary search and autotuned Gather/Scatter can be applied synergistically with these works to achieve significantly high system performance.

\subsubsection*{Deep Learning Compilers.} 
Deep Learning (DL) compilers~\cite{DBLP:journals/corr/abs-1802-04799,10.1145/3575693.3575702,10.1145/3575693.3576933,10.1145/3582016.3582047,10.1145/3133901,won2023unified}
simplify DL programming and automate the hyperparameter search for DL tensor programs, resulting in significant engineering savings. 
However, most DL compilers either optimize dense tensor algebra~\cite{DBLP:journals/corr/abs-1802-04799,10.1145/3575693.3575702,10.1145/3575693.3576933} or sparse tensor algebra with sparsity patterns that do not depend on the input data~\cite{DBLP:journals/corr/abs-1802-04799}.
Since the sparsity pattern of \SpConv networks depends on the particular characteristics of the given 3D point clouds, the tensor programs compiled by these prior DL compilers~\cite{DBLP:journals/corr/abs-1802-04799,10.1145/3575693.3575702,10.1145/3575693.3576933} are still inefficient for point cloud networks. 
To our knowledge, TACO~\cite{10.1145/3133901}, TACO-UCF~\cite{won2023unified}, and SparseTIR~\cite{10.1145/3582016.3582047} are the only DL compilers that optimize sparse tensor algebra by taking into consideration the sparse pattern specified by each particular input data.
However, these DL compilers do not integrate the optimizations proposed in \system.
Thus, \system's four key ideas work synergistically with these DL compilers to provide significant system performance benefits in \SpConv executions.

\subsubsection*{Binary Search Optimizations on GPUs.}
A couple of prior works \cite{8665796,6270834,10.1145/2304576.2304621} explore binary search optimizations on GPUs. 
TriCore \cite{8665796} discusses the cache friendly behavior of na\"ive binary search, when executing lookups with sorted queries. 
However, as discussed in Section \ref{sec:opt-map-step}, building the kernel map in \SpConv by simply executing fully sorted queries in the source array with na\"ive binary search would incur large sorting overheads (Challenge 1), that would offset the cache benefits, and does not explore the optimization on the number of comparisons (Challenge 2).
MergePath \cite{6270834,10.1145/2304576.2304621} improves the computational complexity of na\"ive binary search, however it necessitates a cache-unfriendly binary search process on GPUs, thus causing worse system performance than the na\"ive binary search \cite{8665796}.
We conclude that applying these prior binary search-based schemes in the \Map step would still be inefficient compared to our proposed segmented sorting double-traversed binary search algorithm.

\section{Conclusion}

\system is a highly efficient \SpConv engine that accelerates 3D point cloud networks on GPUs. \system highly utilizes the on-chip GPU memory hierarchy, improves execution parallelism and metadata costs, and reduces unnecessary data accesses and computations on \SpConv executions. 
Our evaluations show that \system significantly outperforms prior state-of-the-art \SpConv engines by \Eval{MinuetE2EAvg} speedup on average at the end-to-end execution, across a wide variety of sparse point cloud networks, datasets, and GPU architectures. We conclude that \system is a novel memory-efficient \SpConv engine tailored for modern GPUs, and hope that our work encourages further comprehensive studies and optimization strategies on point cloud networks and other sparse deep learning networks.

\bibliographystyle{plain}
\bibliography{references}

\end{document}